# Plasmonic polarization sensing of electrostatic superlattice potentials

Shuai Zhang*†[1], Jordan Fonseca†[2], Daniel Bennett[3], Zhiyuan Sun[4,7], Junhe Zhang[1], Ran Jing[1], Suheng Xu[1], Leo He[1], S.L. Moore[1], S.E. Rossi[1], Dmitry Ovchinnikov[2], David Cobden[2], Pablo. Jarillo-Herrero[5], M.M. Fogler[6], Philip Kim[4], Efthimios Kaxiras[3,4], Xiaodong Xu*[2], D.N. Basov*[1]

[1]Department of Physics, Columbia University, New York, NY, 10027, USA

[2]Department of Physics, University of Washington, Seattle, WA, 98195, USA

[3]John A. Paulson School of Engineering and Applied Sciences, Harvard University, Cambridge, Massachusetts 02138, USA

[4]Department of Physics, Harvard University, Cambridge, Massachusetts 02138, USA

[5]Department of Physics, Massachusetts Institute of Technology, Cambridge, MA, USA.

[6]Department of Physics, University of California, San Diego, La Jolla, CA, USA

[7]Current address: Department of Physics, Tsinghua University, Beijing 100084, P. R. China.

†These authors contributed equally to this work.

*Corresponding authors' email addresses:

sz2822@columbia.edu; xuxd@uw.edu; db3056@columbia.edu

**Abstract**

Plasmon polaritons are formed by coupling light with delocalized electrons. The half-light and half-matter nature of plasmon polaritons endows them with unparalleled tunability via a range of parameters, such as dielectric environments and carrier density. Therefore, plasmon polaritons are expected to be tuned when in proximity to polar materials since the carrier density is tuned by an electrostatic potential; conversely, the plasmon polariton response might enable the sensing of polarization. Here, we use infrared nano-imaging and nano-photocurrent measurements to investigate heterostructures composed of graphene and twisted hexagonal boron nitride (t-BN), with alternating polarization in a triangular network of moiré stacking domains. We observe that the carrier density and the corresponding plasmonic response of graphene are modulated by polar domains in t-BN. In addition, we demonstrate that the nanometer-wide domain walls of graphene moirés superlattices, created by the polar domains of t-BN, provide momenta to assist the plasmonic excitations. Furthermore, our studies establish that the plasmon of graphene could function as a delicate sensor for polarization textures. The evolution of polarization textures in t-BN under uniform electric fields is tomographically examined via plasmonic imaging. Strikingly, no noticeable polarization switching is observed under applied electric fields up to 0.23 V/nm, at variance with transport reports. Our nano-images unambiguously reveal that t-BN with triangular domains acts like a ferrielectric, rather than ferroelectric claimed by many previous studies.

**Introduction**

Recently, polar domains have been experimentally observed in artificially stacked van der Waals (vdW) materials, such as transitional metal dichalcogenides (TMD) [1-3] and hBN [4-6]. In these stacked devices, the individual layers are twisted by small angles with reference to the aligned (rhombohedral) commensurate stacking. This particular stacking configuration breaks the inversion symmetry and results in polar domains [7-9]. To illustrate the formation of polar domains, we take hBN as an example. When the B and N atoms in neighboring layers are vertically aligned, forming the AB or BA stacking, the mirror plane between the layers is broken, resulting in charge transfer between layers, manifested by out-of-plane up/down polarizations [10]. For a minimally twisted hBN, a triangular network of nearly uniform AB and BA stacking domains is formed because of lattice relaxation [9,11,12]. The domains are separated by narrow domain walls (DW) and pinned by very localized (energetically unfavorable) AA stacking regions. As a result, the stacking devices form a regular network of triangular polar domains [13,14], where neighboring polar domains possess opposite polarizations. The polarization potential can modulate the carrier density or Fermi energy of proximal materials [15-17], thereby actuating superlattices and creating new functionalities and quantum states [18-21]. For example, by integrating polarization into heterostructures, polarization controlled superconductivity [22], exciton diffusion [17], and non-volatile memory [6,23] were demonstrated.

Moiré polar domains in twisted hBN have attracted intense interest as hBN is vastly used in heterostructures as an excellent encapsulant and gate dielectric, and it is highly desirable to explore the emergent phenomena in heterostructures with polarization from t-BN. Moreover, hBN possesses the largest polarization amplitude, $P \sim 2 \times 10^{12} \, C/m$, of all the discovered interface ferroelectric materials [1] and offers exciting optical properties, including quantum emitters, optical nonlinearity, and hyperbolic phonon polaritons [24]. When t-BN is integrated with graphene field effect transistors, conductivity hysteresis was observed. The hysteresis was attributed to the polarization switching of entire devices [6], even though there is no in-situ and spatially resolved polarization evidence. Whereas polarization switching under electric fields was observed by applying voltages onto the scanning probe tip [1,5,25], the fields therein were inhomogeneous due to the lightning-rod effect and thus cannot be well quantified to enable comparisons with the transport results. Yet, direct imaging of the evolution of the polar domain in t-BN subjected to uniform electric fields has never been achieved, posing a fundamental dilemma for the studies on polar properties of moiré materials. In addition, similar hysteresis was also observed in other heterostructures, such as graphene aligned with hBN, although the reported hysteresis amplitude and mechanism are highly controversial [26,27]. Taken together, to better understand the nature of the polarization and polar domain engineered properties, it is mandatory to map out polarization textures.

Polarizations at exposed surfaces have been examined by Kelvin probe force microscopy (KPFM) [3-5,12,28] and piezoresponse force microscopy(PFM) [29], but these surface-sensitive methods cannot access the buried polarization in practical devices. Recently, the polar domains of twisted WSe$_2$, subject to uniform electric fields generated by graphene electrodes, have been imaged with electron microscopy [2,30,31]. The back scattering electron approach, however, is less sensitive to light elements, for example, boron and nitrogen in hBN [2]. In addition, the filter method for electron diffraction cannot be readily adapted to hBN because one cannot disentangle contributions from t-BN and BN encapsulation [30]. It is worth noting that neither KPFM nor electron microscopy can measure how the polarization influences the adjacent materials, which is critical for fundamental understanding and subsequent applications of these emergent polarizations. Therefore, mapping out the polar domains and their impacts on adjacent materials in a practical device with gating electrodes remains a daunting task. This work aims to explore the plasmon polaritons in polarization-actuated superlattices and to use our established plasmon polariton properties to probe the evolution of polarization textures in response to uniform electric fields.

Plasmon polaritons can be formed by coupling light with delocalized electrons [32,33]. These hybrid collective modes allow significant spatial confinement of electric field [34] and facilitate strong light-matter interactions [35]. Consequently, the plasmon polaritons can be sensitively tuned by a wide range of parameters, such as carrier density and dielectric environments. Thus, plasmon polaritons are promising in the context of quantum sensors [36,37]. Plasmon polaritons are expected to be tuned via electrical polarization because the carrier density can be tuned by polarization potential; conversely, the plasmon polariton response might reveal properties of the underlying polar materials, i.e., sensing polarization with plasmons.

Here we use infrared nano-imaging and nano-photocurrent measurements to investigate the graphene in proximity to t-BN. We observe that the alternating out-of-plane polarizations in the stacking domains of t-BN impose a superlattice potential on the neighboring graphene layer, which spatially modulates carrier density in graphene, thereby forming graphene superlattice. Polarization modulated plasmonic responses in the graphene monolayers are observed. We demonstrate that the electrostatic potential associated with the sharp domain walls of moiré superlattices provides momenta required to excite the plasmonic polaritons. In addition, we establish that the plasmon of graphene could act as a sensor for polarization textures. The evolution of polar domains in t-BN in response to uniform electric fields is examined here for the first time. Strikingly, no noticeable polarization switching is observed under electrical fields up to 0.23 V/nm. Thus, our images unambiguously reveal that the polar domains in t-BN do not prominently respond to an applied electric field. These results suggest that it is more appropriate to classify the polar response of t-BN as a ferrielectric-like material, as opposed to the commonly reported ferroelectric classification.

**Results and discussion**

**Heterostructures and nano-optical experiments**

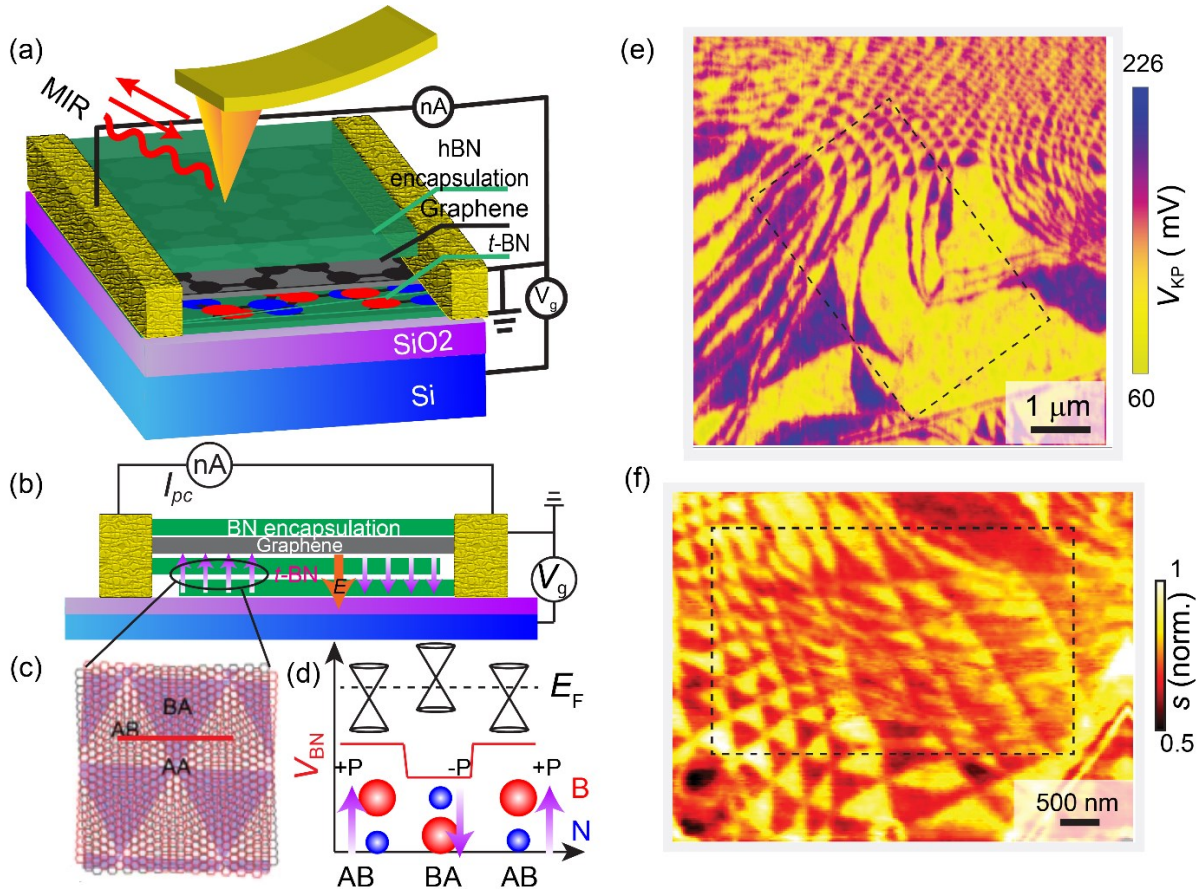

FIG. 1. Moiré polarization investigated by KPFM and scattering-type scanning near-field optical microscope. (a) Schematic of the experimental setup for near-field scattering and nano-photocurrent measurements. Continuous-wave middle infrared (MIR) light is focused on a metallized AFM tip and back-scattered. Arrows near the tip denote the incident and back-scattered light. The scatted light and photocurrent are recorded simultaneously. (b) Schematic of the device structure. Graphene is laid on the marginally twisted hBN (t-BN), which forms a triangular network of out-of-plane polar domains. The alternating polarization directions are denoted by red and black arrows, respectively. The graphene is encapsulated by a thin top hBN bulk (nonpolar) sheet. The electric contacts to graphene are made along its edge. (c) Schematic

of the moiré patterns formed in marginally t-BN. (d) The graphene Fermi energies are modulated by the electrostatic potential of t-BN. (e) Kelvin-probe map showing domains with opposite polarization. The image was recorded during the sample fabrication process, when the graphene and the top encapsulation h-BN had not yet been stacked onto the t-BN. (f) Near-field amplitude image acquired at $V_g = -25$ V after the graphene and top h-BN were laid onto the t-BN. The regions marked by black dashed lines in (e) and (f) are the same. By comparing (e) and (f), we note that the moiré pattern changed during the fabrication while some regions retain a similar texture. The observed change of buried moiré texture implies the need of characterizing finished devices with tomographic imaging. All data were acquired on device 1.

All devices were fabricated using a well-established dry transfer and stacking method [38]. More details on device fabrication can be found in Appendix A. The device structure is illustrated in Figs. 1(a) and 1(b). Monolayer graphene is overlaid on the minimally twisted hBN, possessing alternating out-of-plane polarization in the triangular stacking domains, henceforth denoted by graphene/t-BN. The assembled graphene/t-BN structures were encapsulated by additional nonpolar hBN layers. The entire stacks were finally transferred onto degenerately doped Si wafers with a SiO$_2$ layer thickness of 285 nm, serving as backgate electrodes. The t-BN structures used in this work include twisted bilayer and twisted few-layer structures; these t-BN layers were stacked without intentional rotations, thus forming near-zero rhombohedral stacking with alternating out-of-plane polarization (Fig. 1(c)). The Fermi energy of graphene, $\mu$, is governed by the global Si backgate and the local polarization in t-BN (Fig. 1(d)). The graphene layer is terminated by several edge contacts, for the purpose of simultaneous gateable nano-infrared images, electric transport, as well as nano-photocurrent measurements. The structures of two representative devices are shown in Appendix B.

Nanoscale imaging of polarization engineered plasmon and photocurrent was achieved using the following experimental protocol with a scattering-type scanning near-field optical microscopy (s-SNOM), as illustrated in Fig. 1(a). The metallized tip of an atomic force microscope is illuminated by a focused infrared laser with tunable energy $\omega$. The tip intensifies the incident electric field at the apex, thus allowing probing of electrodynamics of the devices with nanometer scale resolution. The tip was operated in tapping mode, and the scattered light and collected photocurrent were demodulated in proper methods [39]. Thus, we obtained the scattered light and photocurrent arising from local light excitations. More details regarding nano-infrared imaging and nano-photocurrent can be found in Appendix A. The scattered light encodes the local optical conductivity of materials, $\sigma(\mu, \omega)$, and the electric field of plasmon polariton.

**Plasmonic response modulated by electrostatic potential**

The polarization textures of t-BN were confirmed using KPFM measurements during the device fabrication, when the t-BN was exposed without graphene and top hBN encapsulants. The representative KPFM image (Fig. 1(e) and more data in Appendix B), exhibits triangular moiré domains with staggered electrostatic potentials, arising from the alternating polarization of stacking domains. Notably, similar moiré patterns were also visualized by the nano-infrared scattering amplitude image (Fig. 1(f)), which was acquired using s-SNOM on the same finished device after hBN and graphene encapsulation. Thanks to the unique tomographic capability of our nano-infrared imaging, we successfully visualized the polar domains through encapsulants, thereby allowing us to probe the polarization texture in practical devices, which was altered by the final fabrication steps and demonstrated by different moiré textures acquired at the same region in Figs. 1(e) and 1(f). More importantly, the top graphene electrode enables exploring the polarization evolution under uniform electrical fields.

We conjecture that the contrast in the nano-infrared images originates from the plasmonic response of graphene because the Fermi energy of graphene is modulated by the polarization potentials in proximal t-BN layers [6,15]. Now, we systematically survey the plasmon excitations in graphene with moiré superlattices. In nano-infrared experiments, the plasmonic excitation manifests itself in two observables: i) near-field scattering amplitude [40-42]; ii) interference patterns [41,42]. Both observables are analyzed in this study, in order to elucidate the plasmonic behaviors in graphene superlattices.

We first investigate the near-field scattering amplitude evolution with carrier density, which is tuned via the backgate. A series of near-field scattering amplitude images were acquired at various backgate voltages. Representative images are shown in Figs. 2(a-h). In all images, the polar domains are discernable, except at a back gate voltage where the moiré-averaged Fermi energy of the entire device reaches the charge neutrality point (CNP) (Fig. 2(b)). At this particular voltage, the neighboring domains possess the same carrier density but with opposite signs, thus resulting in the same plasmonic response. The near-field signal contrast between the neighboring domains is prominent when the graphene Fermi energy is near the CNP because in the latter regime the carrier density of graphene is dominated by doping from proximal polar domains. In addition, the contrast between neighboring domains can reverse, as evident by comparing domains marked by dashed lines in Figs. 2 (a), (c), and 2(f). The contrast reversal is caused by the non-monotonic evolution of the scattering signal as a function of doping, as we discuss below.

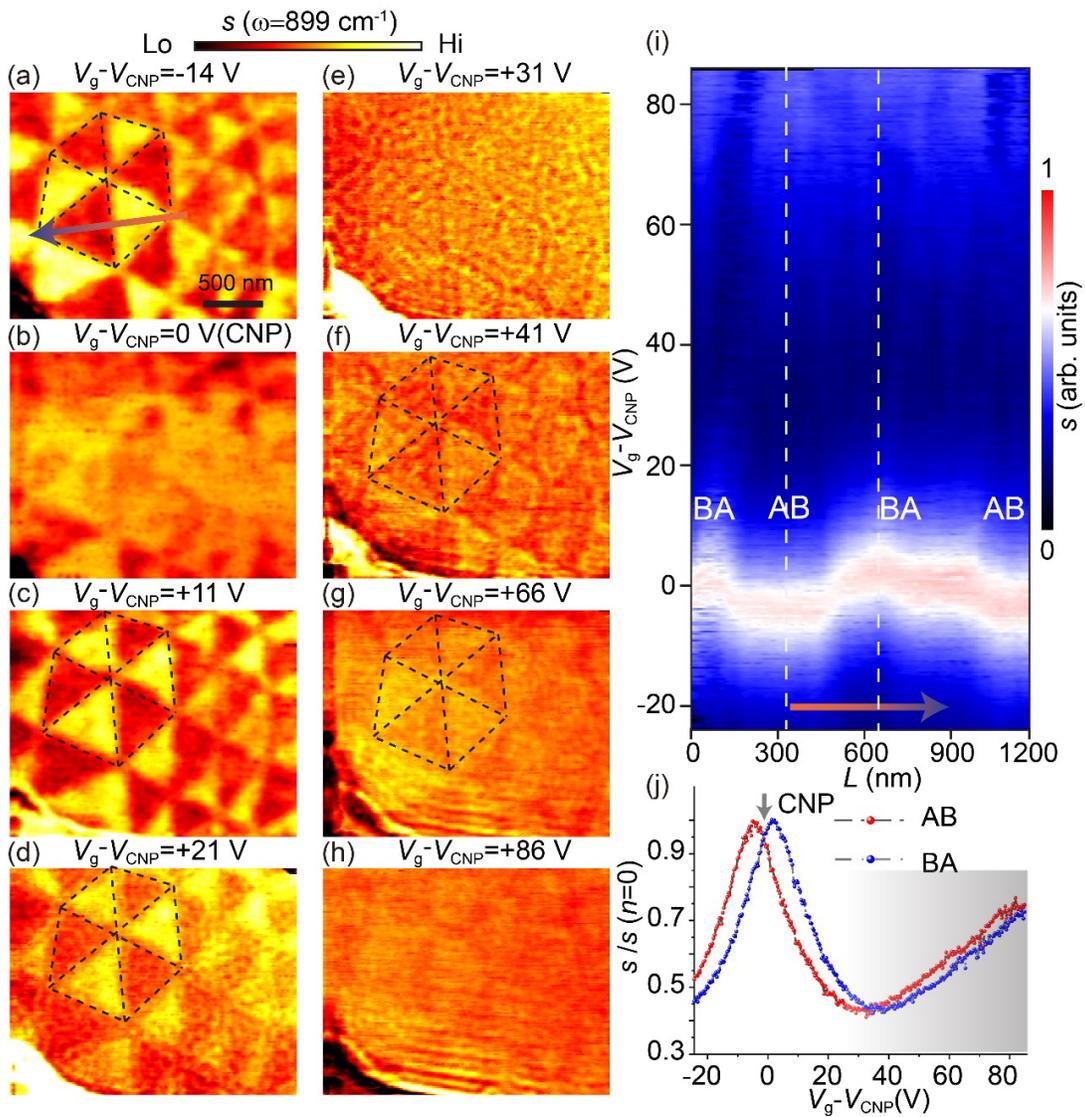

FIG. 2. Plasmon excitation of graphene in proximity to a superlattice potential as a function of electrostatic doping. (a-h) Near-field scattering amplitude $s(\omega)$ of the graphene doped by both electrostatic potential from t-BN and electrostatic gating. The images were acquired at the labeled back gate voltages. Excitation energy $\omega$=899 cm$^{-1}$. A color bar is shown at the top to represent the relative scattering amplitude (that is, low and high). (i) The near-field scattering amplitude as a function of gate-voltage $V_g$ measured from a line trace that crosses several domains. The line trace is denoted by an arrow line in (a). (j) Extracted line cuts highlighted by dash lines in (i). The charge neutrality points corresponding to the AB and BA domains are marked by gray arrows. In the shaded region, graphene is heavily doped, and the tip is coupled with the plasmon polariton modes. All data were acquired on device 1 at 200 K.

Now, we systematically investigate the scattering amplitude evolution arising from gate tunable plasmonic excitations. For this purpose, we repeatedly scanned the tip across domains along the arrow denoted in Fig. 2(a) and gradually swept the back gate voltage; a mapping of the scattering amplitude was obtained, as shown in Fig. 2(i). To clearly illustrate the evolution of near-field scattering amplitude, the line-cuts for the AB and BA domains are extracted from Fig. 2(i) and plotted in Fig. 2(j). The two plots for the AB and BA domains are similar. In both plots, the near-field amplitude reaches a local maximum at CNP. With doping, the amplitude first decreases and then upturns. This trend can be well understood by invoking the Fresnel reflection coefficient $r_p$, which governs the near-field signals. The tip-sample coupling $G$ can be represented by $G = \int r_p w(q) dq$, where $w(q)$ is the weight of tip momentum $q$. The $w(q)$ peaks around $1/a$, where $a$ is the tip curvature radius, and gradually fades when away from $1/a$. So, we focus on $r_p$ at the momenta near the peak of $w(q)$. At low doping regimes, the amplitude of $r_p$, $Amp(r_p)$, decreases with doping, and thus the near-field amplitude decreases. After a doping threshold, the $Amp(r_p)$ increases because the momentum of the dispersive plasmon overlaps with the tip momentum. Then the tip becomes better coupled with the plasmon and generates a strong scattering amplitude (shaded area Fig. 2(j)). Meanwhile, as the momentum offered by the tip can compensate for the momentum mismatch between free light and the plasmon, the tip could launch plasmon polaritons; then, fringes from the interference of propagating plasmon polaritons appear, which can be seen at the bottom of Figs. 2(g) and 2(h). A detailed photon energy dependence of polarization modulated plasmonic response is in Appendix E.

From the plasmonic excitations, we are able to quantify the carrier density induced by the polar domains in t-BN. Nanometer-resolved plasmonic response enables probing the carrier density in each polar domain. From Fig. 2(j), CNP of each domain can be identified, denoted by two arrows. The difference between the two CNPs in terms of backgate voltage is 6.51 V. With the known device capacitor geometry, the voltage difference can be translated into a carrier density difference of $\pm 2.40 \times 10^{11}\ cm^{-2}$ for two neighboring AB and BA domains. The Fermi energy of graphene in two domains is shifted by $\pm 57\ meV$, which is much larger than the calculated 27.5 meV using the reported polarization potential [4]. The CNP analysis and the theoretically calculated polarization potential induced doping are documented in Appendixes F and G, respectively. For device 2, the CNP difference between two neighboring domains is 0.96 V, corresponding to a carrier density of $3.25 \times 10^{10}\ cm^{-2}$. The Fermi energy is shifted by $\pm 21\ meV$, closer to the calculated 14 meV with a domain size of ~2 $um$. By comparing the two devices, we conclude that the CNP of device 1 is at the backgate voltage of ~36 V, indicating substantial extrinsic doping, whereas device 2 with CNP at ~1 V possesses much less extrinsic doping. This less extrinsically doped device shows weaker doping contrast, closer to the predicted value using reported polarization potential. The above results confirm that the graphene carrier density can be tuned by polarization textures, as stated by those transport results [1,6,43]. However, the vastly different doping levels induced by the polarizations of t-BN cannot be simply interpreted by the displacement field originating from a polar material. Our real-space resolved doping measurements suggest that other mechanisms, such as charge transfer between the graphene and defects in h-BN, can intertwine ferroelectric potential and thus enhance the doping difference of neighboring domains. Our results imply that, in order to trace the origin of reported hysteresis in transport experiments, it is critical to read the carrier density in each polar domain directly.

**Domain wall-assisted plasmon polariton excitations**

In scattering amplitude images, besides the contrast between domains discussed above, we also observe "ripples" in the nano-IR intensity within the domains (Figs. 2(d-h)). These ripples emerge at very low carrier density, which is around four times lower than the threshold for the usually reported tip-launched/scattered plasmon polaritons. These ripples do not originate from the inhomogeneity of carrier density, as the homogeneous near-field scattering amplitude of each domain near CNP demonstrates the uniform doping in each domain, as shown by images in Figs. 2(b) and 2(c). More nano-infrared images with ripples can be found in Figs. 9-13. To trace the origin of these ripples in real space images, we perform a Fourier transform of the near-field scattering images, as shown in Figs. 3(a-e). The 2D amplitude spectra reveal several prominent characteristics. First, a hexagonal star is displayed near the center (zero momentum). Second, at higher momenta, two azimuthally symmetric modes, manifesting themselves in rings of $q$ and $2q$, can be observed. The hexagonal star in momentum space originates from the periodic triangular domains. The isotropic wavevectors with a ring-feature are attributed to the ripples in real space, which can be confirmed by the inverse FFT shown in Figs. 3(g) and 3(h). The two wavevectors, $q$ and $2q$, originate from the plasmon polariton interference patterns with periods of wavelength $\lambda$ and half wavelength $\lambda/2$, respectively. These momenta gradually decrease with an increase in the carrier density (Fig. 3(f)). This correlation indicates that the ripples are plasmon polaritons.

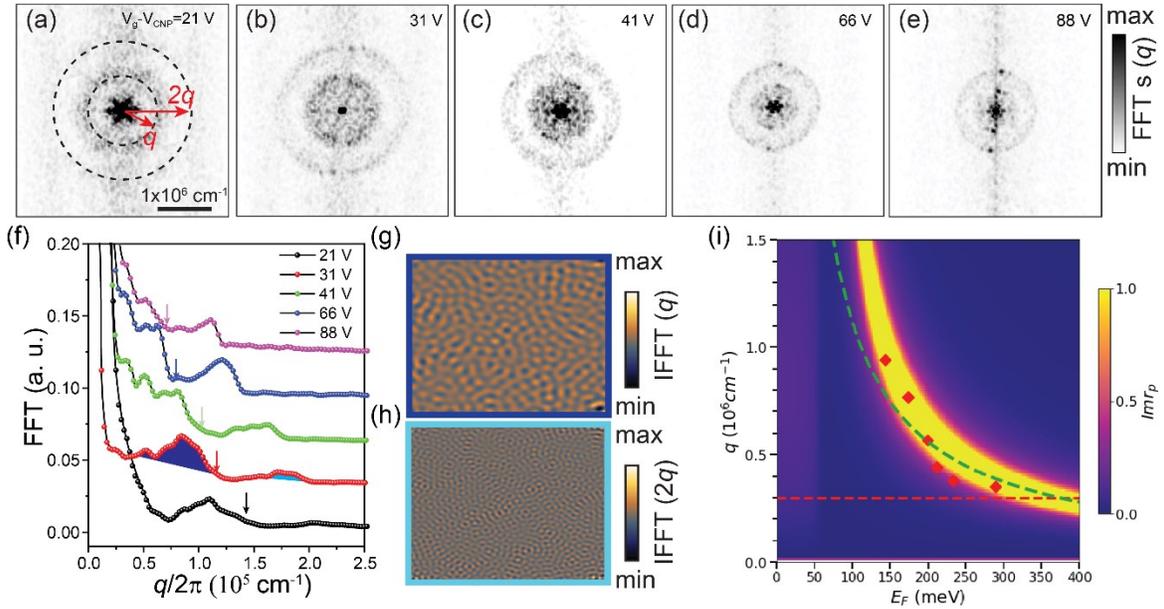

FIG. 3. Superlattices domain wall assisted plasmonic excitations. (a-e) Fourier transformation of images acquired at various backgate voltages. (f) Angle integrated line profiles along the radial direction. (g,h) Inverse Fourier transformation of the momentum q/2 and q, respectively. The amplitude of momentum used in the transformation is marked by the shaded region in (f). (i) Experimental dispersion of ripples (red diamonds) and the calculated dispersion of plasmon polariton dispersion. The dispersion is visualized using the false color map of the imaginary part of reflection coefficient $r_p$, where p indicates p-polarized light. The green dashed line denotes the plasmon dispersion $q \propto 1/\sqrt{E_F}$, with a simplified constant dielectric environment. The red dashed line denotes the momentum at which the tip-sample coupling maximizes.

To unambiguously confirm that the ripples originate from plasmons, we investigate their dispersion. We plot the wave vector $q$ (radius of outer ring in each FFT panel) and the corresponding carrier density in Fig. 3(i). In the same panel, the simulated plasmon dispersion, revealed by $\text{Im}(r_p)$, is also plotted. We can find that the dispersion of the ripples can be well captured by the calculated plasmon polaritons. The well-matched dispersion attests to the conclusion that an emerging plasmon polariton, ripples in real space, is discovered in graphene superlattices, imprinted by polar domains.

Now, we elucidate the mechanism of generating the emerging plasmon polaritons with ripple features in our devices. The plasmon polaritons in two-dimensional electronic systems cannot be excited by far-field illumination. The key notion here is the momentum mismatch: polaritons possess much larger momenmta than those of photons in free space. Therefore, the graphene plasmon polaritons can only be observed in devices with a periodic grating structure [44], an antenna or metallic tip [45], which could provide momentum to satisfy the momentum conservation. Strikingly, here we observed a kind of plasmon with momenta much larger than all the previously observed ones. All the previous experiments used devices with period, $l$, of hundreds of nanometers [44] or tips with radius, $a$, of tens of nanometers [41,42]. Thus, the observed plasmon momenta are limited to $\sim 1/l$ and $1/a$, respectively. However, in the graphene electrostatic superlattices with Fermi energy modulated by polar domains, the sharp domain walls with a width of a few nanometers [11], could provide unprecedentedly large momentum, thereby allowing the excitation of high-momentum plasmon polaritons. The plasmon polaritons launched by the domain walls propagate with wavelength $\lambda$ and interfere with each other, forming ripples. In this regime, the tip serves as a sensor for the local electric field of domain wall-assisted plasmon polaritons. By gradually increasing the doping, the plasmon momentum decreases and matches with the momentum that the tip provides, marked by a red dashed line in Fig. 3(i). As a result, the tip can launch plasmon polaritons, which are reflected by the boundaries and result in a round trip, thus forming a period of $\lambda/2$ (ref. [42]). From Figs. 3(a-f), we can see that with an increase in doping, the two polaritons compete: the intensity of $q$ declines as that of $2q$ grows.

The capability of exciting plasmon with the aid of moiré domain walls provides new opportunities for exploiting the strong coupling of light and matter. For the domain wall-assisted plasmon polariton, at low doping, the tip only works as a sensor for the electric field and is used to read out the plasmon polaritons. Upon shedding light on the moiré superlattices, the plasmon polaritons are formed on the entire moiré device, thanks to the momentum imparted by domain walls. We also note that domain walls in the present work do not show high reflection to polariton because the optical conductivity contrast at the walls is subtle. By further increasing the polarization induced doping difference between neighboring domains, such as via reducing the distance between graphene and backgate, the increased conductivity at the domain boundary will efficiently reflect polaritons. Thus, photonic crystals for plasmons could be formed, like in twisted bilayer graphene [46].

**Evolution of polarization amplitude and domain shape with electric field**

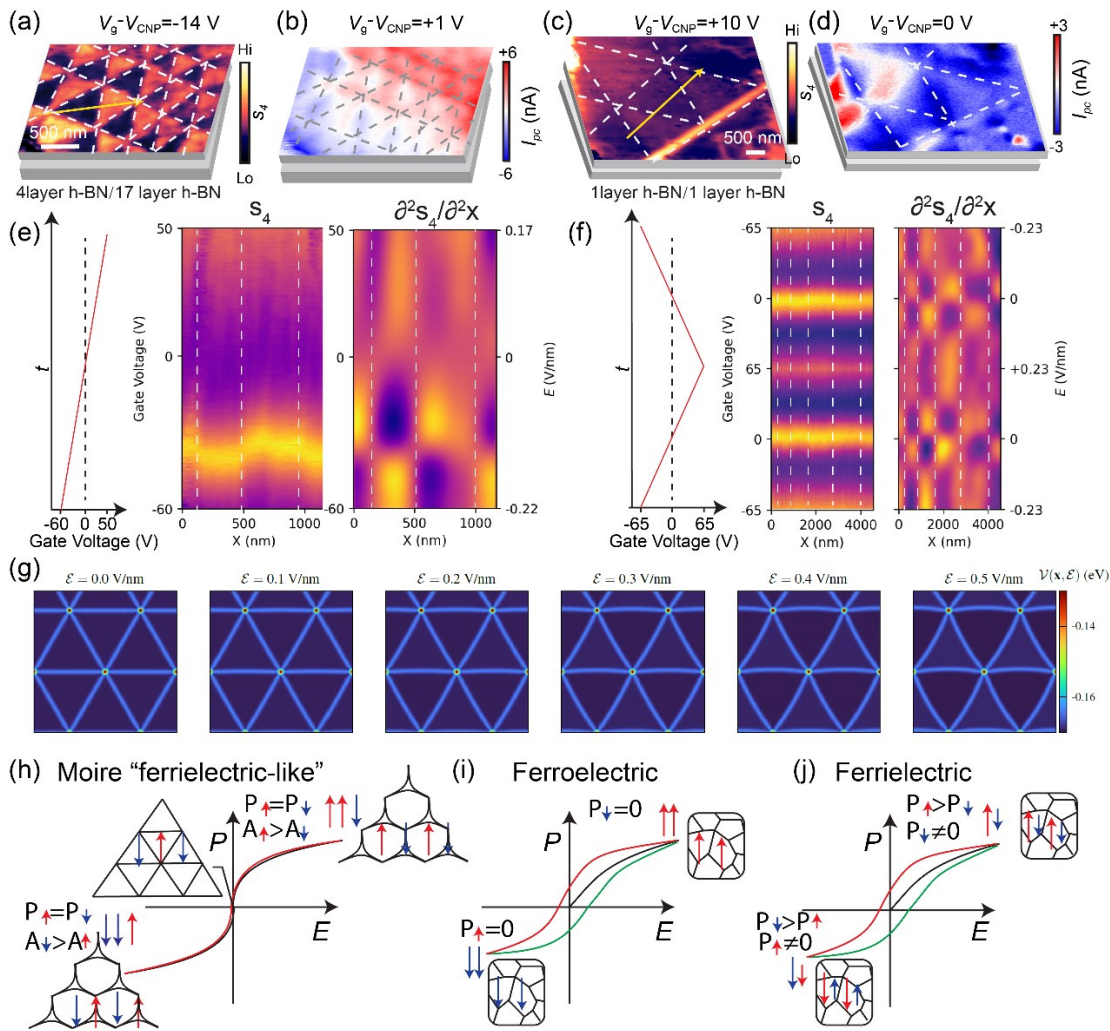

FIG. 4. Polarization evolution with uniform electric fields. (a and c) Images of the near-field scattering amplitude, $s_4$, acquired on twisted few layer and twisted bilayer BN, respectively. The thickness of hBN is marked. (b and d) Images of the nano-photocurrent acquired in the regions of (a) and (c), respectively. (e and f) The near-field scattering amplitude (middle panel), $s_4$, and its 2$^{nd}$ derivative with respective to spatial position (right panel), $\frac{\partial^2 s_4}{\partial x^2}$, plotted as a function of gate-voltage $V_g$ and measured along line traces indicated by yellow lines in (a) and (c), respectively. The white dashed lines mark the domain walls. The left panels show the sweep of the backgate voltage and the corresponding electric field as a function of time. (g) Stacking energies in t-BN bilayer after lattice relaxation, as a function of position, for a twist angle of $\theta$=0.1°. When an electric field is applied, the AB and BA domains relax unevenly: one slightly expands; the other correspondingly shrinks. (h, i and j) Evolution of the total polarizations, averaged over domains, as a function of electric field in moiré polar materials such as t-BN (h), in prototype ferroelectric materials (i), and prototypical ferrielectric materials (j). Illustrations of the typical domain structures for strong positive and negative electric fields are sketched in each case.

We have demonstrated that polarization textures can be visualized and quantified by monitoring the plasmonic responses. With this newly established plasmonic polarization sensor, we revisited the notion of the domain shape evolutions under electric fields [1,6,25,30]. We investigated both twisted few-layer BN and twisted bilayer BN. The domains can be identified in both nano-IR images and photocurrent images (Figs 4(a-d)). More details on nano-photocurrent are in Appendix D. To visualize the polarization evolution, we repeatedly scanned the tip along a line across several polar domains, marked by the arrow lines in Figs. 4(a) and 4(c), and swept the gating voltage (left panels of Figs. 4(e) and 4(f)) simultaneously. Thus, a mapping of near-field scattering amplitude as a function of position and backgate voltage was formed, as shown in the middle panels of Figs. 4(e) and 4(f). The swept backgate voltages have two separate consequences: i) graphene carrier density is tuned by the voltages due to geometric capacitance, where the graphene and backgate silicon function as two electrodes of the capacitor; ii) t-BN is subjected to the resultant electric fields, formed between graphene and backgate electrode. The former doping effect tunes the plasmonic response and thereby prompts the evolution of the scattering amplitude. The latter field effect may reshape the domains, as polarization switching is realized by the bending of domain walls: domains with polarization aligned with the field may be expected to grow, and the domains with polarization antialigned with the field correspondingly shrink [7,13,14].

Now, we examine the response of t-BN to uniform electric fields, tunable via the backgate. The polar domain induced local plasmonic contrasts are obscured by the backgate induced global plasmonic response. To remove the background and thus highlight the polar domain induced contrasts, we take the second derivative of the near-field amplitude with respect to the tip position, $\frac{\partial^2 s_4}{\partial x_{tip}^2}$ (ref. [47]), as shown in the right panels of Figs. 4 (e) and 4(f). The locations of the domain walls correspond to contrast change along the tip position $x$ and are marked by white dashed lines. In the $\frac{\partial^2 s_4}{\partial x_{tip}^2}$ mapping, the contrast between the domains can be clearly identified. Strikingly, the domain walls do not show noticeable bending under uniform fields up to 0.23 V/nm. With higher fields, devices are exposed to the risk of dielectric breakdown in hBN or SiO$_2$. Namely, the total polarizations, averaged over domains, are far from saturation at the practically reachable electric fields. Steady domain structures/polarization textures were observed in both twisted few-layer BN and twisted bilayer BN. This result, the first nanoscale-resolved polarization response of t-BN to uniform electric fields, questions the assumed coercive field of less than 0.1 V/nm in previous transport measurements [6].

To gain insights into the physics uncovered by our observations, we performed lattice relaxation calculations for t-BN, which provide a theoretical estimation of the domain wall bending in response to an applied electric field, see Fig. 4(g) (for more details in Appendix I). At zero field, the t-BN shows a regular network of triangular stacking domains with alternating polarization, averaging to zero over a moiré period, antiferroelectric-like polarization configurations (Fig. 4(h)). With increasing field strength, the AB and BA domains relax unevenly, and the domain walls separating them bend. However, this bending of the domain wall is negligible even at fields up to 0.5 V/nm, consistent with our experimental observations. Both our experimental images and calculations unambiguously show that the average polarization per t-BN moiré period does not saturate for realistic field strengths. Under large electric fields, beyond the experimentally reachable regimes, the opposite polarizations cannot be totally canceled, resulting in ferrielectric-like features: AB and BA domains with two opposite polarizations coexist but have unequal areas, thus resulting in finite net polarization (Figs. 4(h) and 4(j)). Our observations are very similar to that of t-WSe$_2$ revealed by electron microscopy [30], where the triangular domains with opposite polarization only slightly reshape at a field larger than 0.2 V/nm [30]. Since the BN lattice is more rigid than WSe$_2$, the domain reshaping via bending of the domain wall is expected to be even more difficult, as confirmed by our data.

In addition to polarization switching, hysteresis in response to electric fields is another hallmark of ferroelectric material. However, for twisted bilayers, the theoretically predicted domain growth and shrinking under electric fields do not exhibit hysteresis [13]. Although a narrow second-order hysteresis has been experimentally observed in WSe$_2$, this feature is due to inevitable defects in stacking samples, such as the pinning of domain walls by bubbles as the electric field is applied and removed [30]. In short, neither a saturation of total polarization nor hysteresis were observed, in a defect-free t-BN with triangular domains under realistic electric fields. We remark that the near-field amplitude contrasts between the domains reverse during the backgate voltage (electric field) is swept (along vertical axes in Figs. 4 (e) and (f)). The contrast reversal is due to the nonmonotonic evolution of the scattering signal with doping, rather than a switching of the polarization. This nonmonotonicity and contrast reversal can be clearly seen by comparing the near-field amplitude line profiles acquired on two domains, shown in Fig. 2(j).

We note that for untwisted samples, and for samples with very small twist angles, i.e. in the monodomain limit, a switching of polarization can be achieved via a global relative sliding between the layers (van der Waals sliding), rather than the bending of domain walls [5,30]. We also stress that our results and the reported reshaping of domains through voltage-biased tip reconcile with each other because the nonuniform electric fields therein are usually much higher than what can be reached in a practical device. In addition, the fields applied via tip were usually underestimated because the local electric field enhancement arising from the sharp tip geometry was improperly neglected [25]. Furthermore, in these tip biased experiments, the ferroelectric materials were exposed; encapsulating moiré superlattice with graphene or other materials will further stabilize the domains since there are interactions between moiré interface and the encapsulating flakes. In our experiments, the same devices without polarization switches show robust hysteresis (Appendix C). Our results beckon reassessment and examination of the reported hysteresis loops.

**Summary and outlook**

In summary, we demonstrate that alternating moiré polar domains result in an electrostatic superlattice potential, which, via proximity, modulates the graphene carrier density. The nano-infrared images reveal plasmonic excitation in graphene superlattices. Notably, the domain wall can assist high momentum plasmonic excitations. This novel plasmonic excitation, provides a new opportunity for strong light-matter coupling. In future studies, we could explore the light-engineered emerging quantum states under the strong coupling regime. Furthermore, by imaging the plasmonic response at the nanometer scale, we examine the polarization evolution in t-BN integrated field effect transistors. Our observed unaltered domain textures and polarizations under electric fields indicate that the observed transport hysteresis in moiré materials cannot be entirely attributed to polarization switching. Our established approach of tomographically visualizing the polarization and their evolution can also be used to understand the polarization mechanism and dynamics of the emerging ferroelectric materials, such as $WTe_2$, and the highly debatable transport hysteresis in other vdW heterostructures, such as the graphene aligned with hBN [26,27]. Furthermore, the moiré potential modulated carrier density and plasmons observed here are expected to be extended to moiré exciton physics [48] and correlated electron phenomena [49].

**Data availability**

The raw data in the current study are available from the corresponding authors upon request.

**Code availability**

The code used for the analysis and simulations in the current study is available from the corresponding authors on request.

**ACKNOWLEDGMENTS**

Nano-imaging research at Columbia is supported by DOE-BES Grant No. DE-SC0018426. Research at Columbia and University of Washington on moiré superlattices is supported as part of Programmable Quantum Materials, an Energy Frontier Research Center funded by the U.S. Department of Energy (DOE), Office of Science, Basic Energy Sciences (BES), under Award No. DESC0019443. D.N.B. is a Moore Investigator in Quantum Materials EPIQS GBMF9455.

S.Z., D.B., P.J.H, P.K., E.K. and D.N.B. acknowledge the support from the ARO MURI programme (W911NF-21-2-0147).

**APPENDIX A: METHOD**

**Fabricate of device with a twisted few layer BN.** Exfoliated flakes of hBN were transferred using the standard dry transfer method with a PDMS/PC stamp. First, thin hBN was picked up and transferred onto an adjacent flake to form the moiré-patterned substrate. Both flakes were transferred onto a silicon wafer with a 285 nm layer of dry-processed $SiO_2$. PC was washed off in chloroform, and the hBN substrate was vacuum annealed at 400 ºC for 2 hours to help remove bubbles and improve layer adhesion. The desired FE-moiré region of the device was AFM cleaned in contact mode on a Bruker Icon AFM with an OTESPA-R3 tip. KPFM was used to identify regions of the substrate with out-of-plane polarized relaxed moiré domains. A second PDMS/PC stamp was used to pick up a thin piece of hBN and graphene and transfer it onto the target region of the substrate as identified with KPFM. After washing off the PC film, EBL and reactive ion etching were used to remove everything around the target sample area. 2D edge contact was made to the graphene edges with a second round of EBL and subsequent evaporation of Au (70 nm) on a sticking layer (7 nm Cr). The finished device was AFM cleaned again to remove PMMA residue that could interfere with the SNOM tip.

**Fabrication of device with twisted bilayer BN fabrication.** Monolayer hBN was torn-and-stacked to form a long-period relaxed moiré using a 30 nm piece of hBN on a stamp that was modified to allow the PC film to be removed from the stamp using thermal-release tape. After the transfer, the stamp was heated to activate the thermal release tape, and the PC film was carefully removed, inverted, and placed sample-side down onto a flat, clean PDMS puck on a glass slide. The PC film was removed by dipping the slide and PDMS puck in NMP for ~2 seconds roughly 15 times. Note that NMP is selected because it does not cause the PDMS puck to swell, unlike acetone or chloroform. The PDMS puck was washed with IPA, and the flakes were deposited from the PDMS puck onto a silicon wafer with a 285 nm layer of dry-processed $SiO_2$. The moiré substrate was vacuum annealed at 250 ℃ for 2 hours, then AFM cleaned in contact mode on a Bruker Icon AFM with an OTESPA-R3 tip. KPFM was used to identify regions of the substrate with out-of-plane polarized relaxed moiré domains. A second PDMS/PC stamp was used to pick up a thin piece of hBN and graphene and transfer it onto the target region of the substrate as identified with KPFM. After washing off the PC film, EBL and reactive ion etching were used to remove everything around the target sample area. 2D edge contact was made to the graphene edges with a second round of EBL and subsequent evaporation of Au (70 nm) on a sticking layer (7 nm Cr). The finished device was AFM cleaned again to remove PMMA residue that could interfere with the SNOM tip.

**Nano-infrared scattering experiments.** The nano-infrared scattering experiments were performed using a home-built cryogenic scattering-type scanning near-field optical microscope (s-SNOM) housed in an ultra-high vacuum chamber with a base pressure of $\sim 7 \times 10^{-11}$ torr. The s-SNOM is based on a tapping-mode atomic force microscope (AFM). The tapping frequency and amplitude of the AFM are about 285 kHz and 70 nm, respectively.

The s-SNOM works by scattering tightly focused light from a sharp AFM tip. The spatial resolution of the s-SNOM is predominantly set by the tip radius of curvature $a$ (~20 nm in our experiment) and therefore allows us to resolve optical and optoelectronic properties at a length scale below that of the polar domains. The laser source is a wavelength tunable quantum cascade laser (QCL) from Daylight Solutions. Photon energy of 860 ~ 920 $cm^{-1}$ was used to avoid phonon resonances from substrate h-BN and $SiO_2$. The laser beam was focused onto the metallized AFM tip using a parabolic mirror with a 12-mm focal length. The back-scattered light was registered by a mercury cadmium telluride (MCT) detector and demodulated following a pseudoheterodyne scheme [39]. The signal was demodulated at the *nth* harmonic of the tapping frequency, yielding background-free images. To eliminate the far-field background, we chose $n$ = 3 and 4 in this work.

**Nano-photocurrent experiments.** The nano-photocurrent measurements were performed simultaneously in the nano-IR imaging experiment. The laser power was set to be ~ 5 mW. The current was measured using a current amplifier with a gain setting of $10^7$ and a corresponding bandwidth larger than 1 MHz. In order to isolate the photocurrent generated by the near fields underneath the tip, the photocurrent was sent to a lock-in amplifier and demodulated at the *nth* harmonic of the tip tapping frequency. In this work, $n$ = 2 was used.

### APPENDIX B: DEVICE CHARACTERIZATION

All the devices were characterized by KPFM before the graphene and BN encapsulation were stacked. KPFM was used to identify regions of the twisted BN with out-of-plane polarized relaxed moiré domains. Two representative devices are shown in Fig. 5 and Fig. 6.

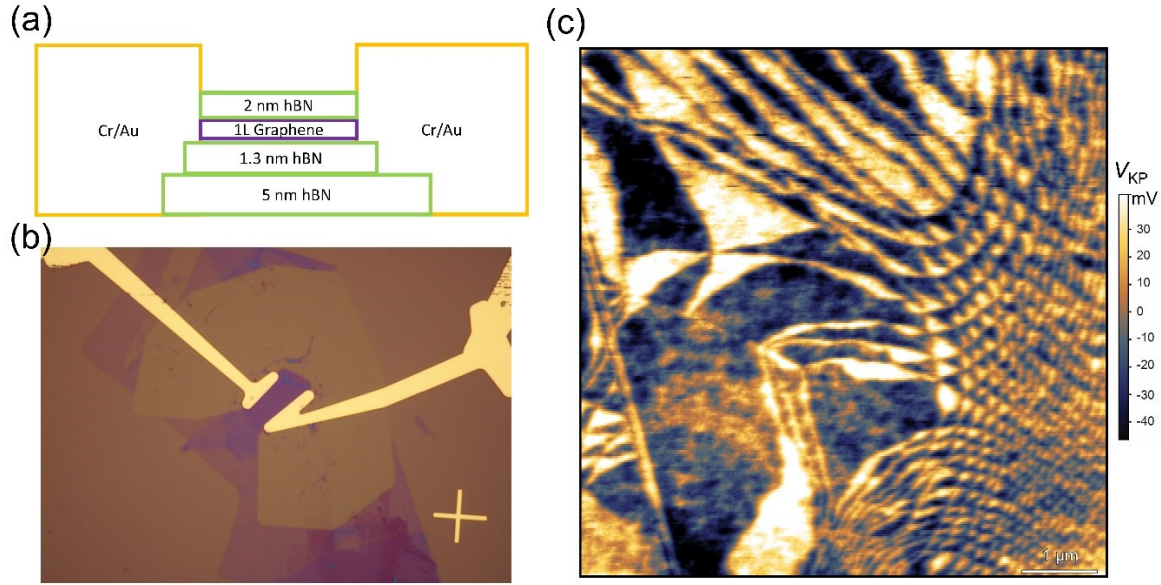

FIG. 5. Device 1. (a) Schematic of device 1. (b) Optical microscope image of device 1. (c) The KPFM image of t-BN measured during the device stacking process.

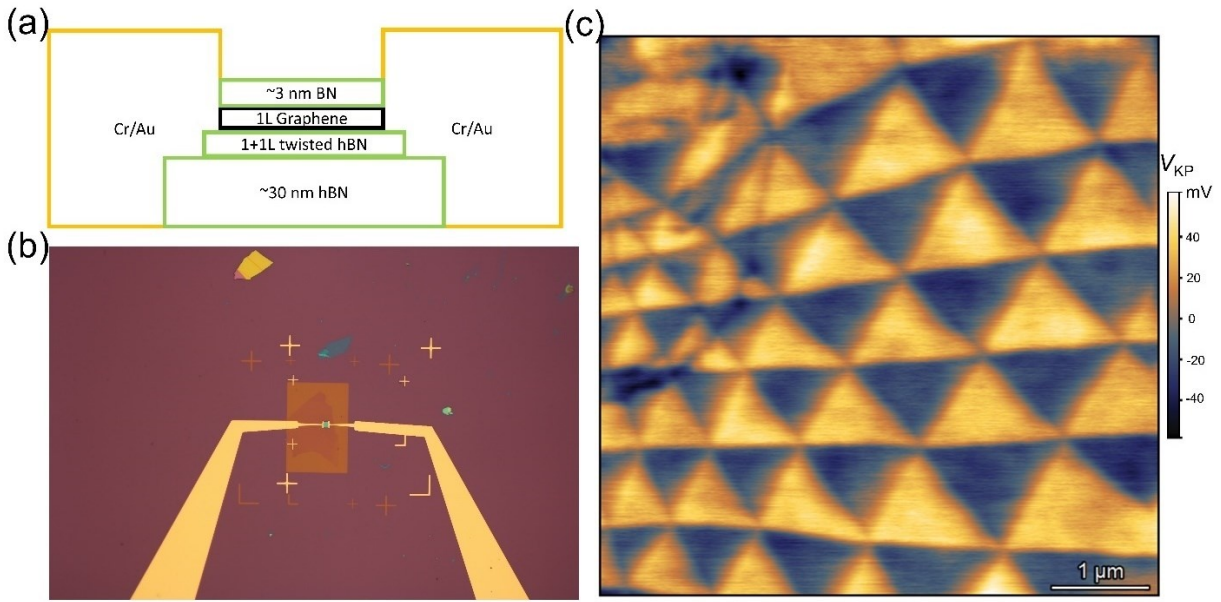

FIG. 6. Device 2. (a) Schematic of device 2. (b) Optical microscope image of device 2. (c) The KPFM image of t-BN measured during the device stacking process.

**APPENDIX C: ELECTRICAL TRANSPORT MEASUREMENTS**

The transport measurement was performed in situ during the near-field infrared (IR) experiments. In the nano-IR imaging experiment, we did not observe clear domain reshaping, which is concomitant with the polarization switching, but prominent and well-reproducible hysteresis was observed in the transport experiment, as shown in Fig. 7. Therefore, the nano-IR and in-situ transport results indicate that the hysteresis is not from the polarization switching.

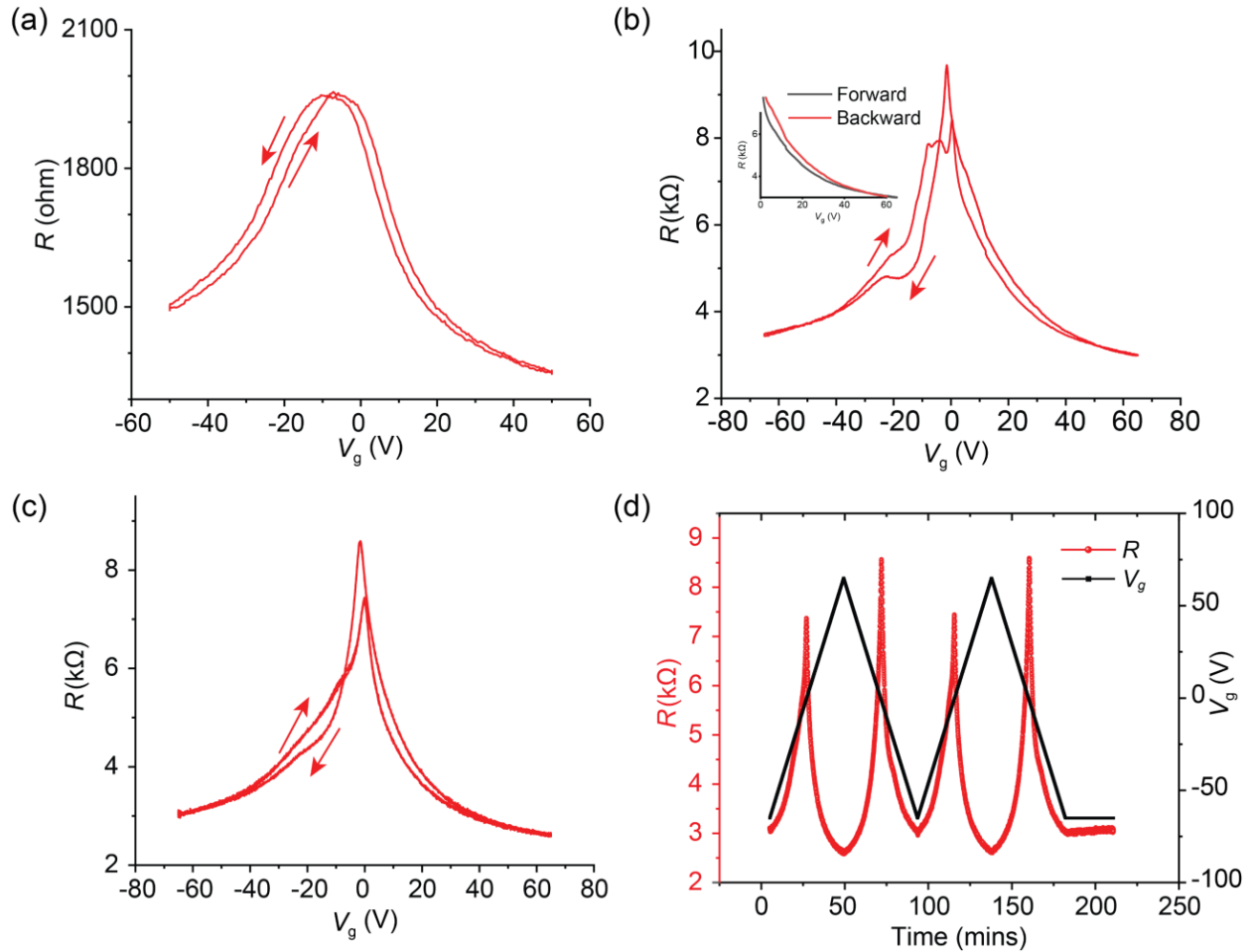

FIG. 7. In-situ electric transport measurements. (a) Resistance of graphene as a function of backgate for device 1 at 295 K. (b) (c) Resistance of graphene as a function of backgate for device 2 at 78 K and 295 K, respectively. The inset of b shows the sharp kink reported in the literature. (d) The resistance of graphene when the backgate is swept forward and backward for two cycles.

**APPENDIX D: NANO-PHOTOCURRENT**

The moiré pattern can be visualized by nano-photocurrent, as shown in Figs. 8(b-f). Photocurrent in graphene is dominated by the photo-thermoelectric effect [15]. The current collected by the contacts can be expressed as:

$$I_{pc} = \int_\Omega \sigma S \nabla \delta T \cdot \nabla \psi d^2 \mathbf{r} = \int_{\partial\Omega} \sigma S \delta T \nabla \psi \cdot d\hat{\mathbf{n}} - \int_\Omega \sigma \delta T \nabla \psi \cdot \nabla S d^2 \mathbf{r}$$

where $\sigma$ is the DC conductivity, $S$ is the Seebeck coefficient, $\delta T$ is the increased electron temperature by light illumination, and $\psi$ is an auxiliary field. The first term on the right-hand side of the equation denotes the photocurrent formed at the contact edges, which results in a photocurrent background. This background fades away when the tip is moved away from the contact edges. The decay length is determined by the electron cooling length. The second term shows that the photocurrent can be detected if a device has a nonzero Seebeck coefficient gradient, $\nabla S$, along the auxiliary field $\nabla \psi$.

As the graphene Fermi energy is modulated by polar domains, the Seebeck coefficient, as a function of Fermi energy, forms a triangular checkerboard pattern. Therefore, a nonzero Seebeck coefficient gradient is formed at the domain boundaries and results in photocurrent. Near the CNP, the Seebeck coefficient gradient between neighboring domains reaches a maximum, and thus the maximum photocurrent forms (Figs. 8 (h) and 8(i)). It is worth noting that blurry domain texture in the nano-photocurrent images is rooted in the electron cooling length.

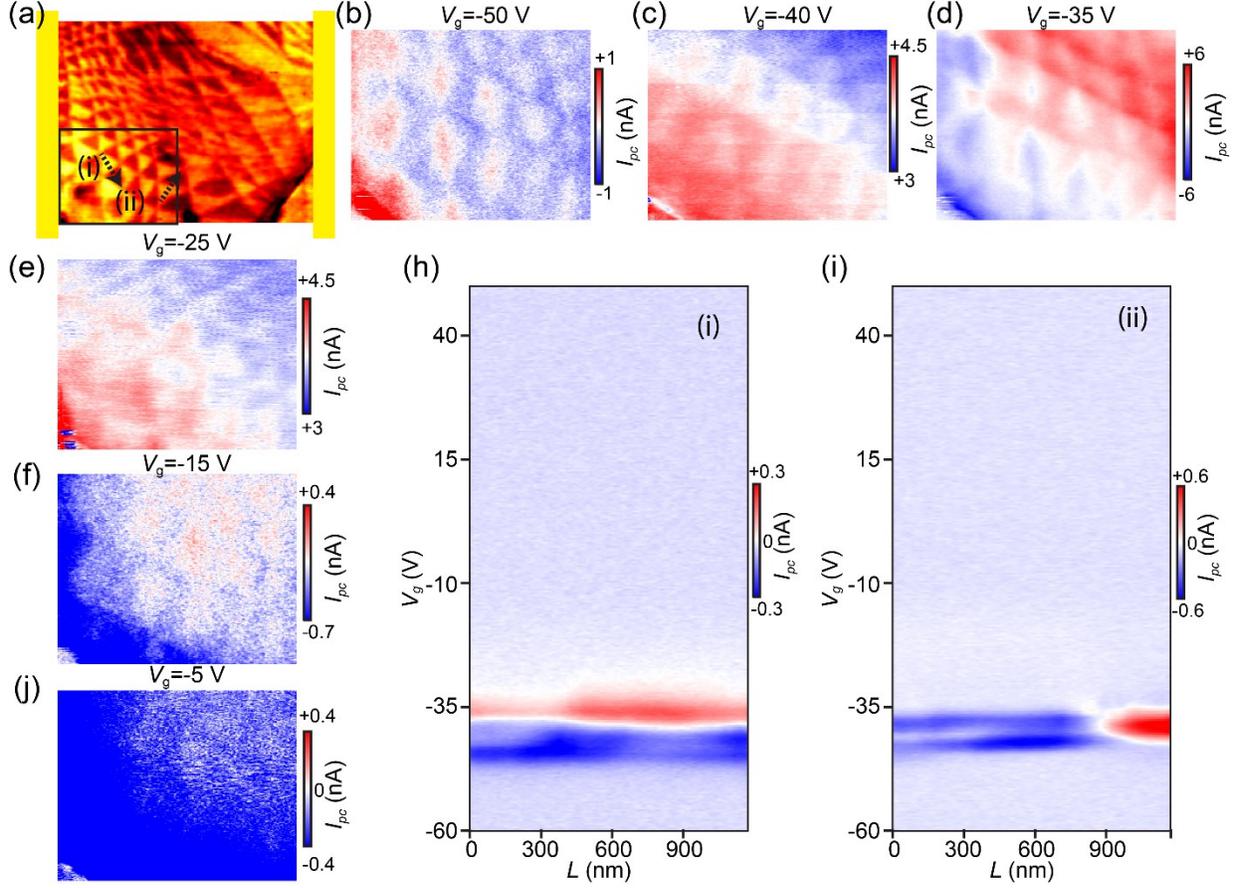

FIG. 8. Nano-imaging of polar domain engineered photo-thermoelectric current. (a) Near-field scattering amplitude image at $V_g$ =-45 V, $\omega$=899 cm$^{-1}$. The gold annotations illustrate the relative positions of the contacts used for measuring photocurrent. (b-j) Gate voltage dependence of the photocurrent ($I_{PC}$) at $\omega$=899 cm$^{-1}$. The back gate voltage is labeled above each panel. (h, i) Photocurrent as a function of gate-voltage $V_g$ measured from a line trace that crosses several domains. The line traces are denoted by arrow lines (i) and (ii) in a, respectively. All data were acquired on device 1. Photocurrent images (b-j) were acquired at 300 K and all other data were acquired on device 1 at 200 K.

**APPENDIX E: PHOTON ENERGY DEPENDENCE OF ELECTRIC POLARIZATION ENGINEERED PLASMONIC EXCITATION.**

We investigated the photon energy dependence of the polarization modulated plasmonic response by measuring nano-IR imaging at various phonon energies (Figs. 9-15). Under various photon energies, infrared spectral features are similar to those described in the main text. To clearly illustrate this dependence, we plot the extracted scattering amplitude line profiles in Fig.

15. When graphene is being doped away from CNP, the scattering amplitude first decreases to a minimum and then increases. This increased scattering amplitude arises from the better coupling between the plasmon modes and the tip. When the photon energies increase, this coupling occurs at higher carrier doping. This finding further confirms that the observed scattering amplitude contrast originates from plasmon excitations because the graphene plasmon energy $\omega_p$, and the carrier density $n$ obey the scaling rule, $\omega_p \propto \sqrt[4]{n}$.

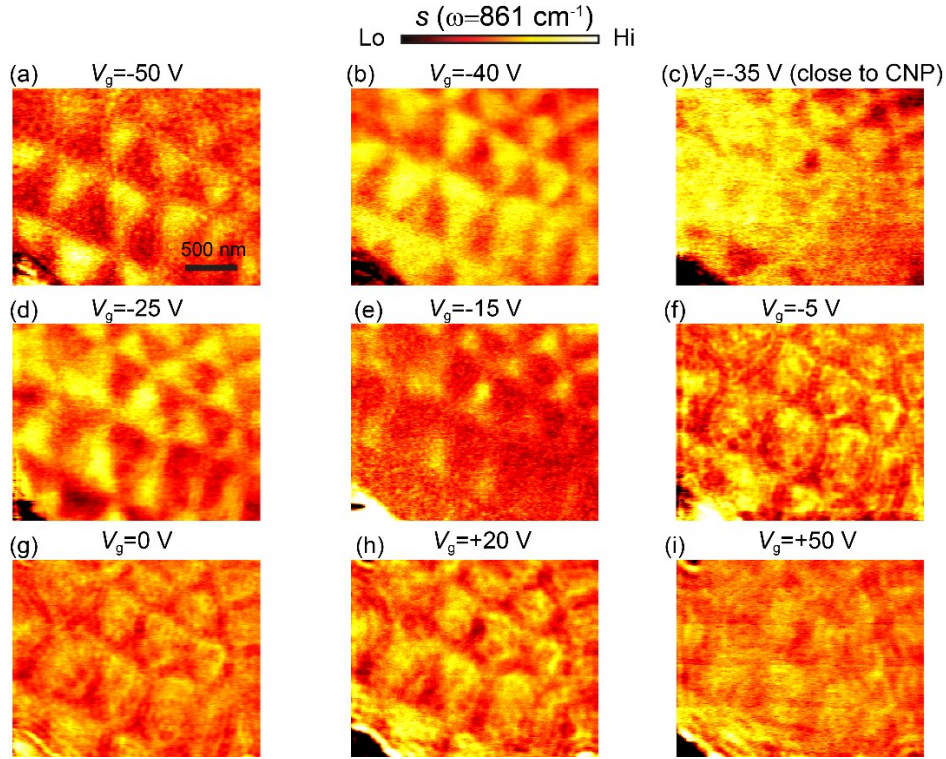

FIG. 9. Plasmonic excitation of graphene in the polar moiré cell as a function of electrostatic doping. (a-i) Nanoscale infrared images of near-field scattering amplitude $s(\omega)$ of the graphene doped by both electric polarization and electrostatic gating. The images were acquired at the labelled back gate voltages. Excitation energy $\omega$=861 cm$^{-1}$ and $T = 300\ K$.

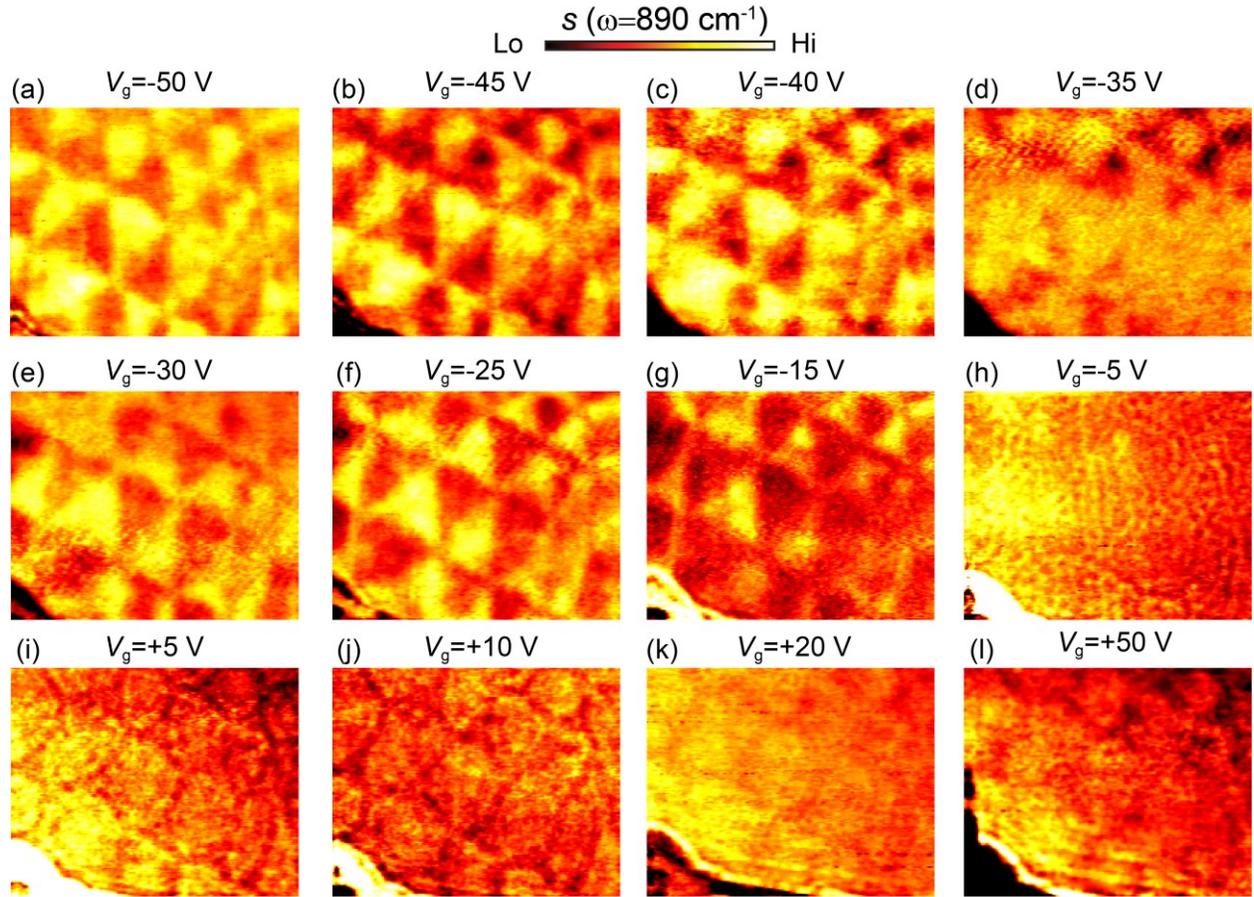

FIG. 10. Plasmonic excitation of graphene in the polar moiré cell as a function of electrostatic doping. (a-l) Nanoscale infrared images of near-field scattering amplitude $s(\omega)$ of the graphene doped by both electric polarization and electrostatic gating. The images were acquired at the labelled back gate voltages. Excitation energy $\omega$=890 cm$^{-1}$ and $T = 300\ K$.

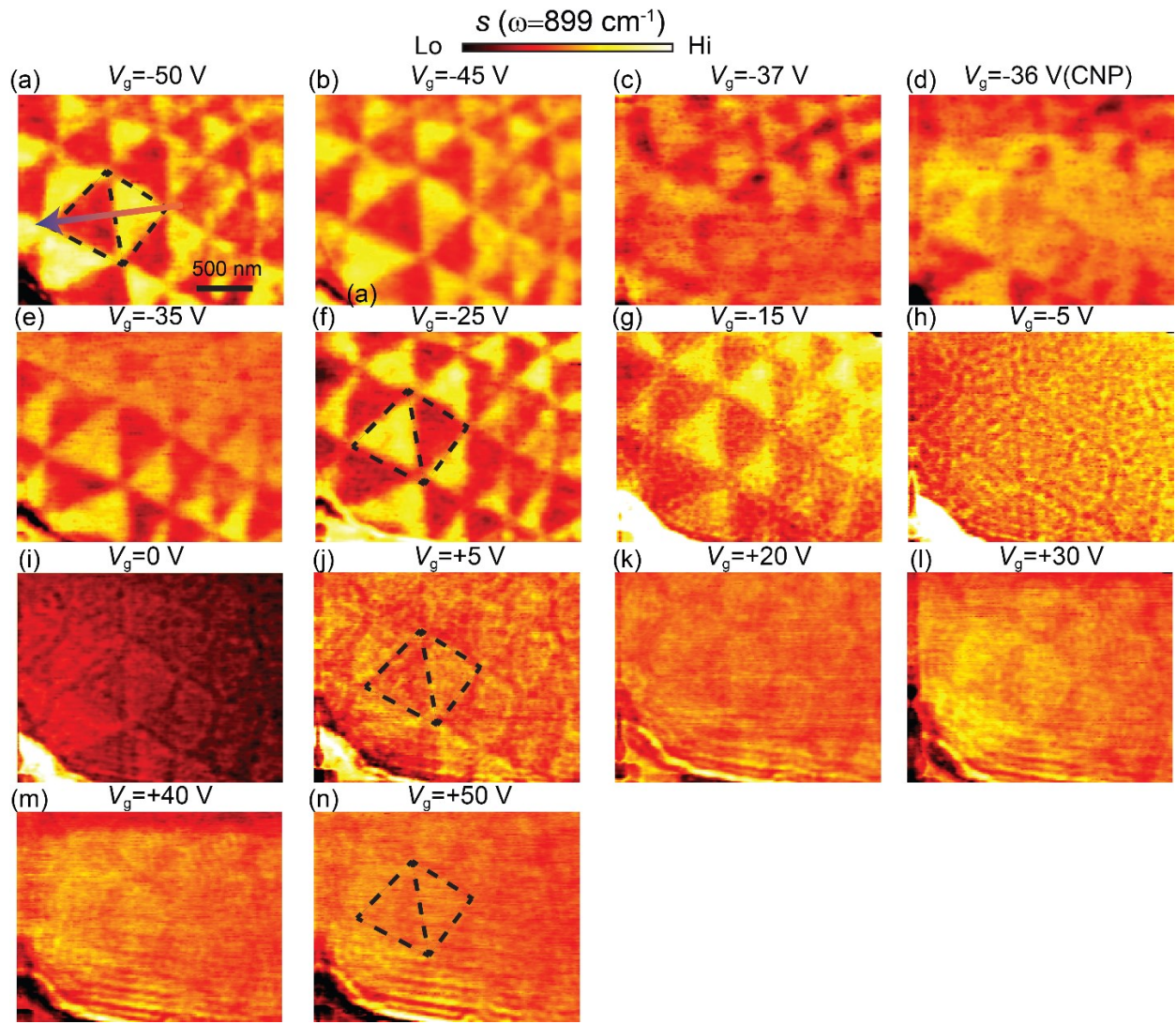

FIG. 11. Plasmonic excitation of graphene in the polar moiré cell as a function of electrostatic doping. (a-n) Nanoscale infrared images of near-field scattering amplitude $s(\omega)$ of the graphene doped by both polarization and electrostatic gating. The images were acquired at the labelled back gate voltages. Excitation energy $\omega$=899 cm$^{-1}$ and $T = 200\ K$.

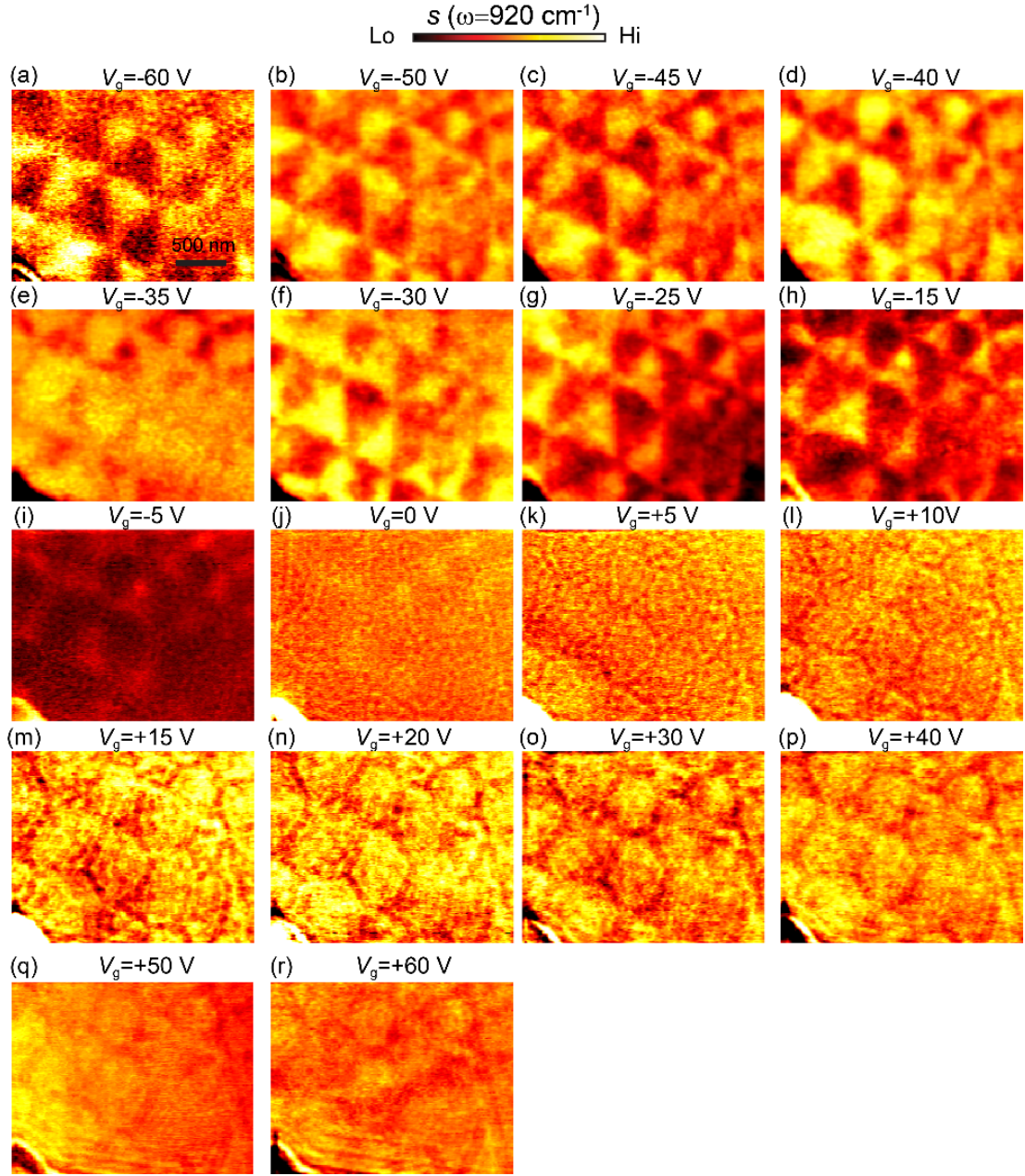

FIG. 12. Plasmonic excitation of graphene in the polar moiré cell as a function of electrostatic doping. (a-r) Nanoscale infrared images of near-field scattering amplitude $s(\omega)$ of the graphene doped by both electric polarization and electrostatic gating. The images were acquired at the labelled back gate voltages. Excitation energy $\omega$=920 cm$^{-1}$ and $T = 300\ K$.

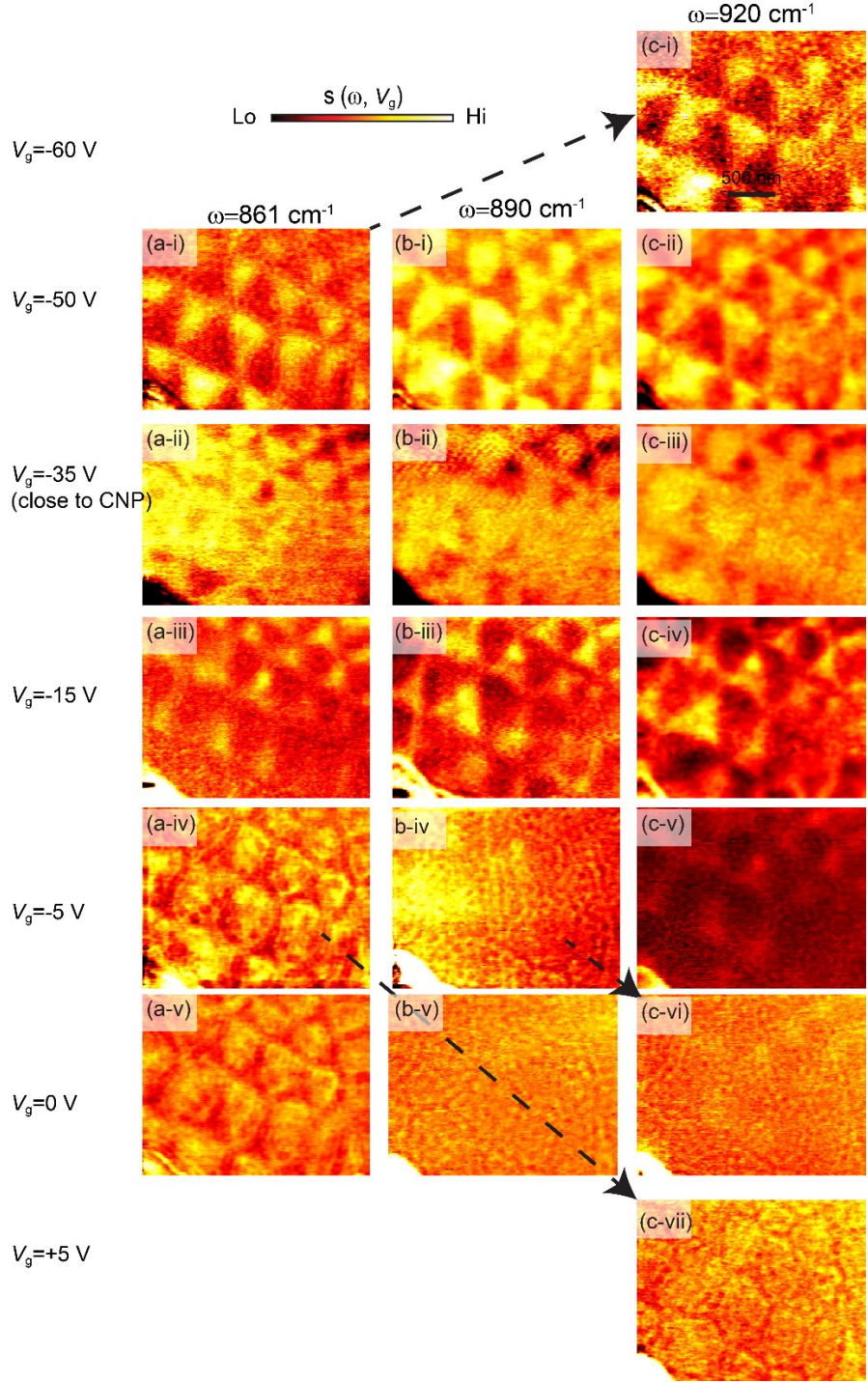

FIG. 13. Comparison of plasmonic excitations with different photon energies. (a-c) Nanoscale infrared images of near-field scattering amplitude $s(\omega, V_g)$ of the graphene at various back gate voltages and photon energies. The back gate voltages are labelled on the left of the images and the photon energies on the top. All data were acquired at $T = 300\,K$. The small dots or ripples in images (ai), (ci), (biv) and (cvi) are from the plasmon interference. For large photon energies, the plasmon interference emerges at higher carrier density, in accord with $\omega \propto n^{1/4}$. Note: the average charge neutrality point of the sample is at $V_g = \sim -36\,V$.

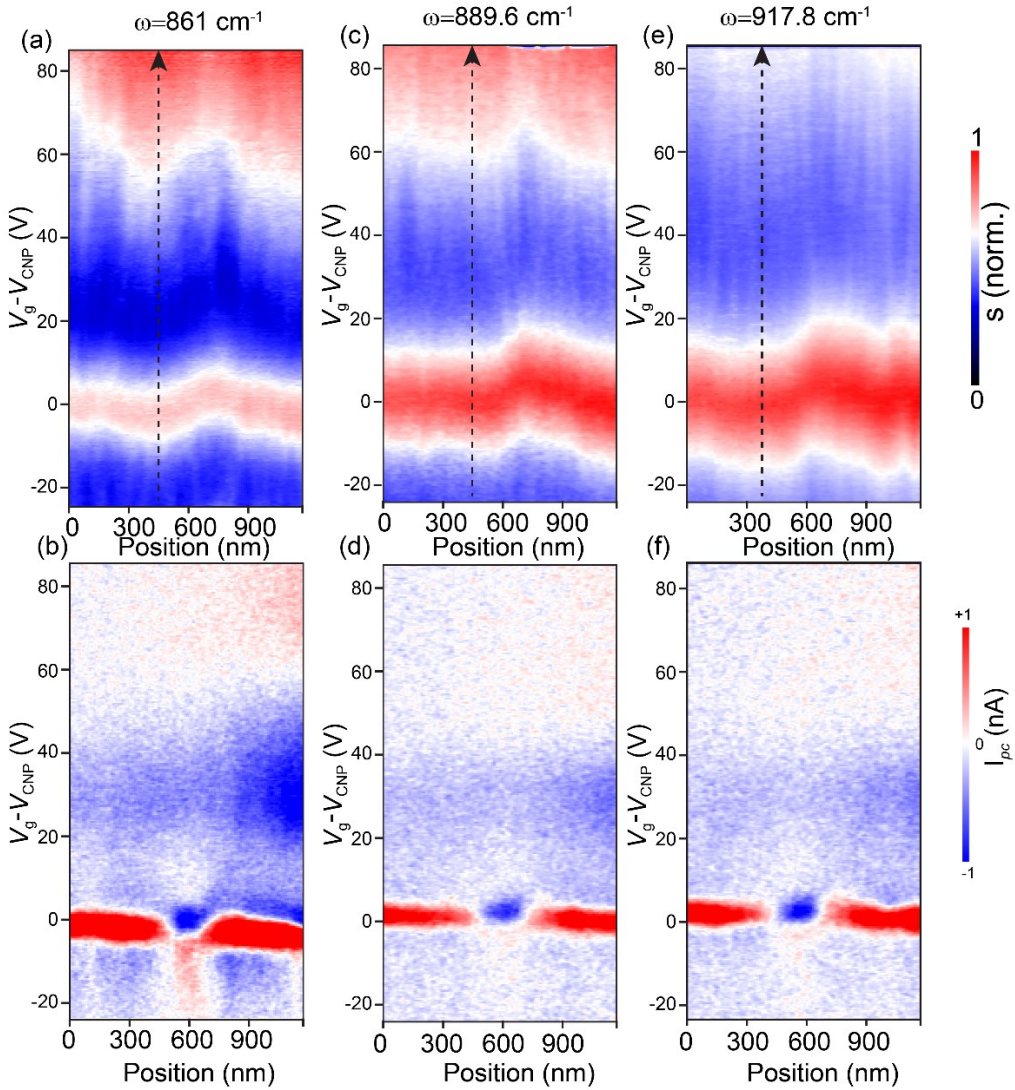

FIG. 14. Comparison of plasmonic excitation and photocurrent at different photon energies. (a, c, and e) The fourth harmonic of the near-field scattering amplitude, $s_4$, plotted as a function of gate-voltage $V_g$ and measured for a line trace that crosses ferroelectric domains. (b, d, and f) The photocurrent, plotted as a function of gate-voltage $V_g$ and measured for a line trace that crosses ferroelectric domains. The photocurrent peaks near CNP.

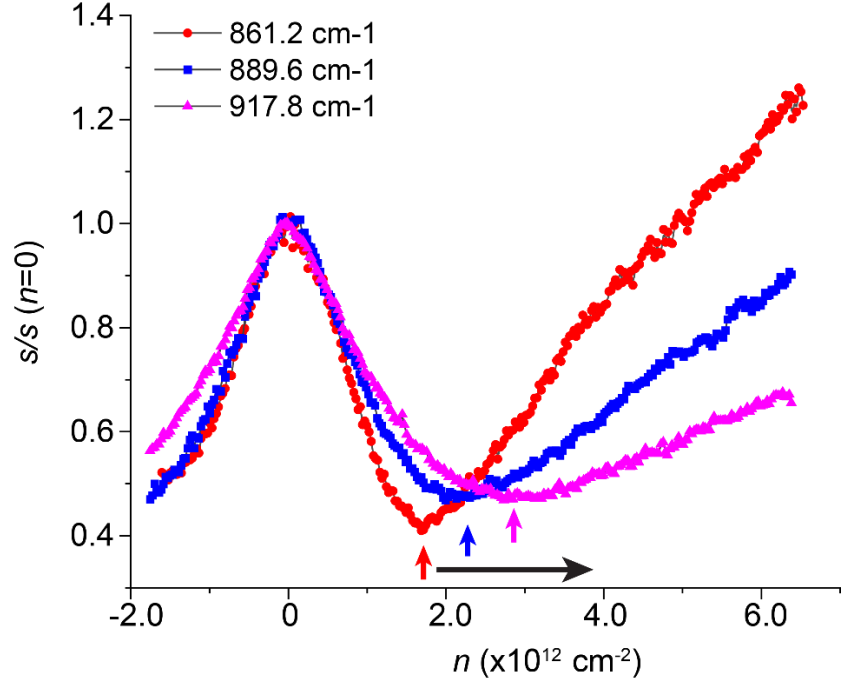

FIG. 15. Near-field amplitude as a function of carrier density at three representative photon energies. The data at the three photon energies are extracted from Fig. 14, in which the line profiles are marked by black dashed lines. The increase in the near-field amplitude above the carrier density $n = 2 \times 10^{12}\ cm^{-2}$ arises from the resonance of tip momentum with the plasmon dispersion. For large photon energies, the plasmon resonance emerges at higher carrier density, indicated by three arrows, in accord with $\omega \propto n^{1/4}$. Note: the arrows indicate the carrier densities at which the plasmon resonance will occur. The gradual increase in the near-field scattering amplitude beyond the denoted carrier density originates from better coupling between the AFM tip and the plasmon.

## APPENDIX F: CNP SHIFT ANALYSIS

Near-field scattering amplitudes $s_4$ as a function of gate-voltage $V_g$ is extracted in the AB and BA domains respectively and interpolated for finer resolution. The Pearson product-moment correlation between the two amplitude profiles is evaluated across a range of shift voltages. The shift voltage at which maximum correlation occurs is defined as the CNP shift voltage. The CNP shift analysis of two representative devices is shown in Fig. 16.

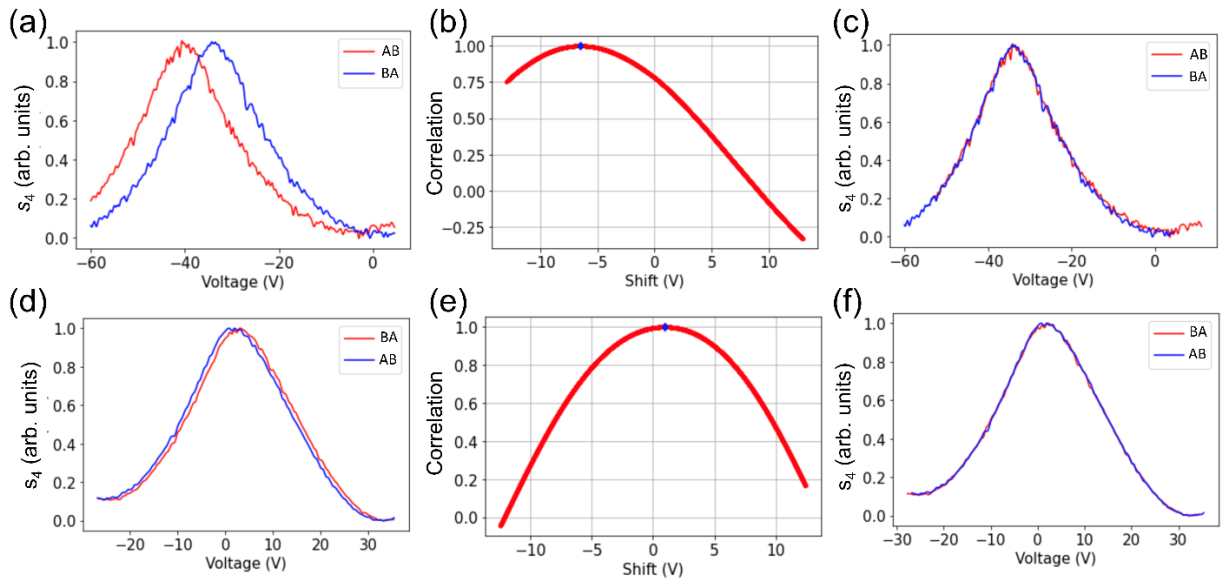

FIG. 16. The CNP shift voltage analysis. (a-c) Polarization induced CNP shift of neighboring domains in device 1. (d-f) Polarization induced CNP shift of neighboring domains in device 2. (a) and (d) show the extracted raw data. (b) and (c) are correlation analysis of the two profiles. The shift of CNP corresponds to the maximum correlation. (c) and (f) show that the two profiles overlap well after being horizontally shifted by the values obtained in correlation analysis.

# APPENDIX G: ESTIMATION OF POLARIZATION POTENTIAL INDUCED DOPING

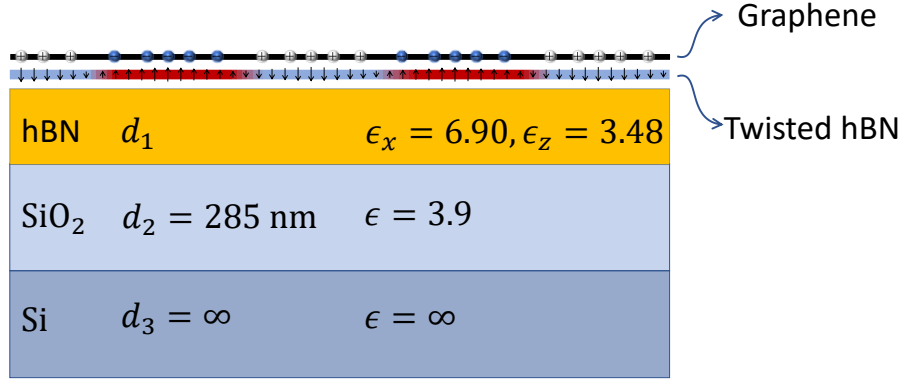

FIG. 17. Schematic of the device structure. The arrow lines on twisted BN denote polarization. The screening from the substrate is encoded in the near field reflection coefficient $R_p(q, \omega = 0)$ (Ref. [45]). For high dielectric screening, $R_p$ is close to $-1$.

The electrical potential immediately above a 2D plane made of 'polar' moiré superlattices can be well approximated by

$$\phi_f(\mathbf{r}) = \phi_0 \begin{cases} 1 & \mathbf{r} \text{ in AB domain} \\ -1 & \mathbf{r} \text{ in BA domain} \end{cases}$$

Due to lattice relaxation, the domain walls are much thinner than the domain period. Away from the 2D plane, this potential decays with a decay length at the order of the moiré period. The periodic ferroelectric potential causes doping of graphene placed parallel to the plane of moiré superlattices. The resulting local Fermi energy $\mu(\mathbf{r})$ of graphene is determined by

$$\mu + \phi_f + \phi[\rho(\mu)] = V_G$$

where $\phi(\mathbf{r}) = \int d\mathbf{r}' \frac{1}{\epsilon} \left( \frac{1}{|\mathbf{r}-\mathbf{r}'|} - \frac{1}{\sqrt{4d^2+|\mathbf{r}-\mathbf{r}'|^2}} \right) \rho(\mathbf{r}')$ is the screening electrical potential due to the doped charge on graphene, $\rho = \frac{1}{\pi} \frac{\mu^2}{v_F^2 \hbar^2} \text{Sign}[\mu]$ is the local charge density, $V_G$ is the gate voltage and $d$ is the distance of graphene to the gate. In the simple case of a stripe moiré lattice, the potential has the analytical form:

$$\phi_f(x) = \phi_0 \sum_n \frac{4}{\pi} \frac{1}{2n+1} e^{-(2n+1)\frac{2\pi}{L}h} \sin\left((2n+1)\frac{2\pi}{L}x\right) = \phi_0 \frac{2}{\pi} \text{ArcTan}\left( \frac{\sin\frac{2\pi}{L}x}{\sinh\frac{2\pi}{L}h} \right)$$

where $L$ is the strip period and $h$ is the distance away from the 2D plane.

### 1. Twisted hBN with finite domain size

For the device shown in Fig. 17, where graphene is on top, the electrical potential generated at the height of graphene by the moiré polarization is $\phi = \phi_f(1 - R_p)$. The screened Coulomb potential of charge on graphene is $V = (1 + R_p)V_q$. The resulting screening charge in graphene can be represented as a doping chemical potential satisfying,

$$(1 + R_p)\frac{\mu^2}{\varepsilon_L} + \mu = \phi_f(1 - R_p), \qquad \varepsilon_L = \hbar\frac{\pi}{L}v_F\frac{\hbar v_F}{e^2},$$

$$\mu = \frac{\varepsilon_L}{2(1 + R_p)}\left(-1 + \sqrt{1 + \frac{4(1 - R_p^2)\phi_f}{\varepsilon_L}}\right) \to \begin{cases} (1 - R_p)\phi_f & \text{if } \phi_f \ll \frac{\varepsilon_L}{4(1 - R_p^2)} \\ \sqrt{\frac{1 - R_p}{1 + R_p}\varepsilon_L\phi_f} & \text{if } \phi_f \gg \frac{\varepsilon_L}{4(1 - R_p^2)} \end{cases}$$

This is the situation of the twisted hBN device. Given a moiré period of 500 nm and an $SiO_2$ thickness of 285 nm such that the screening from the gate is not important, one has $R_p \approx \frac{1-\epsilon}{1+\epsilon} = -0.6$ and $\varepsilon_L = 1.9$ meV. Therefore, the ferroelectric potential $\phi_f \approx 100$ meV is much larger than $\frac{\varepsilon_L}{4(1-R_p^2)}$ and the doping level is about $\sqrt{\frac{1-R_p}{1+R_p}\varepsilon_L\phi_f} \approx 27.5$ meV.

Experimentally, the gate voltage shift between AB and BA domains is $6.51\,V$, meaning the gate voltage needed to cancel the doping in a domain is $3.26\,V$. From the capacitance $\frac{1}{C_g} = 4\pi e\left(\frac{d_1}{\epsilon_z} + \frac{d_2}{\epsilon}\right)$ of the device, it corresponds to a doping density of $n = 2.40 \times 10^{11}$ cm$^{-2}$, considering the thickness $d_1 = 6.3$ nm and out-of-plane dielectric $\epsilon_z = 3.48$ for hBN, and thickness $d_2 = 285$ nm and dielectric $\epsilon = 3.9$ for $SiO_2$. This density corresponds to a doping level of 57 meV, twice as large as the theoretical estimate. We note that if $R_p = -0.92$, meaning $\epsilon_{SiO_2} = 24$, the predicted doping would match the experiment.

### 2. Single domain with infinite size

Note that as $L \to \infty$, one has $R_p \to -e^{-2\pi\frac{2d}{L}}$ neglecting the dielectric screening of the spacer (SiO$_2$). As a result, the doping chemical potential becomes $\sqrt{\frac{1-R_p}{1+R_p}\varepsilon_L \phi_f} \to \sqrt{\frac{2L}{4\pi d}\hbar\frac{\pi}{L}v_F\frac{\hbar v_F}{e^2}\phi_f} = \sqrt{\hbar v_F \frac{1}{2d}\frac{\hbar v_F}{e^2}\phi_f}$ which simply says the doping charge is equal to that induced by $\phi$ on a capacitor with capacitance $C = \frac{4\pi}{ed}$. This theoretical result is consistent with the one used in the previous transport data analysis [6], although the models are not the same.

**APPENDIX H: DOMAIN SHAPE (POLARIZATION) AS A FUNCTION OF ELECTRIC FIELD**

Here we show the domain texture under various gate voltage and the corresponding electrical field. The domain texture does not evolve with electric field, regardless of the domain size. The data shown here and data Fig.4 unambiguously confirm that the polarization does not switch under practically reachable electric fields.

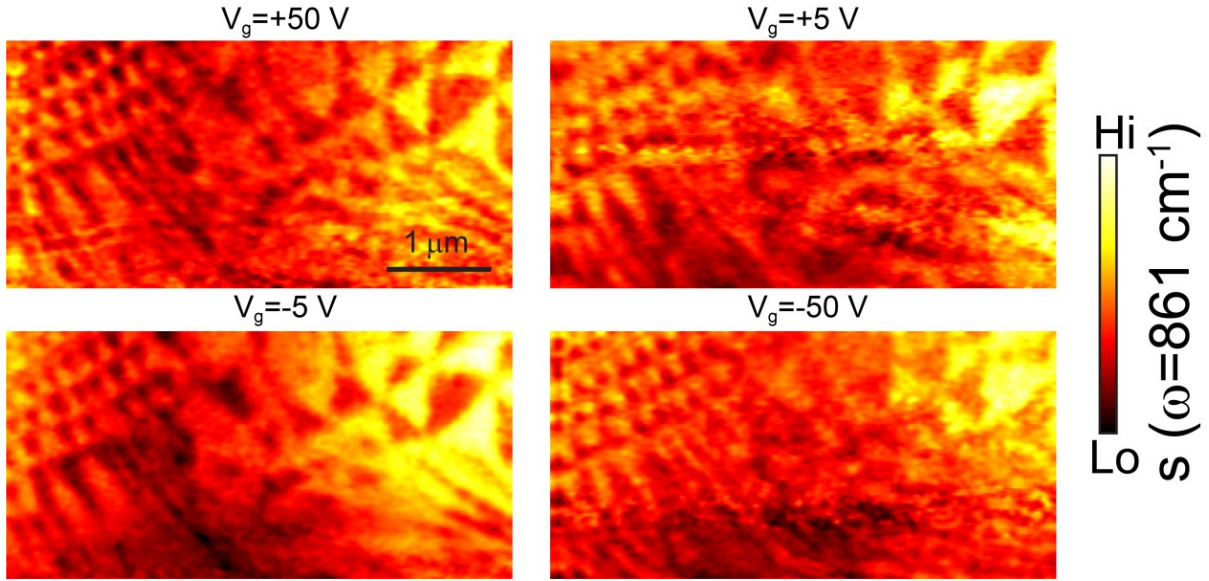

FIG. 18. Domain (polarization) evolution under electric field. Nanoscale infrared images of near-field scattering amplitude $s(\omega)$ of the graphene doped by both ferroelectric and electrostatic gating. The images were acquired at the labelled back gate voltages. Excitation energy $\omega$=861 cm$^{-1}$ and $T = 300\ K$.

**APPENDIX I: LATTICE RELAXATION CALCULATIONS**

Lattice relaxation calculations were performed following the methodology in Refs. [13,14] for t-BN with a twist angle of $\theta=0.1°$ with respect to the ideal rhombohedral stacking (perfectly aligned layers). The total energy of t-BN is given by

$$V_{tot} = \int \mathcal{V}_{tot}(x+u(x))dx$$

$$\mathcal{V}_{tot}(x) = \mathcal{V}_{stack}(x)+\mathcal{V}_{elastic}(x)+\mathcal{V}_{elec}(x)$$

Where $\mathcal{V}_{tot}(\mathbf{x})$ is the total energy density as a function of relative stacking $\mathbf{x}$ between the layers, and $\mathbf{u(x)}$ is a displacement field, which describes the relaxation of the bilayer from its rigid twisted configuration. The integration is performed in "configuration space", in terms of the relative stackings between the layers, all of which are contained in a single primitive cell of bilayer hBN. The total energy density is given as a sum of three independent terms. The stacking energy, $\mathcal{V}_{stack}(\mathbf{x})$,

$$\mathcal{V}_{stack}(\mathbf{x})=\sum_i \mathcal{V}^e_n \phi^e_n(\mathbf{x})+\mathcal{V}^o_n \phi^o_n(\mathbf{x})$$

describes the van der Waals or cohesive energy between the layers. It is written as a Fourier expansion using even and odd $\mathcal{C}_3$ symmetric basis functions $\phi^{e/o}_n$:

$$\phi^e_1 = \cos(2\pi x) + \cos(2\pi y) + \cos(2\pi(x+y))$$

$$\phi^e_2 = \cos(2\pi(x-y)) + \cos(2\pi(2x+y)) + \cos(2\pi(x+2y))$$

$$\phi^e_3 = \cos(4\pi x) + \cos(4\pi y) + \cos(4\pi(x+y))$$

$$\phi^o_1 = \sin(2\pi x) + \sin(2\pi y) - \sin(2\pi(x+y))$$

$$\phi^o_2 = \sin(2\pi(y-x)) + \sin(2\pi(2x+y)) - \sin(2\pi(x+2y))$$

$$\phi^o_3 = \sin(4\pi x) + \sin(4\pi y) - \sin(4\pi(x+y))$$

where x and y are fractions of the primitive lattice vectors of hBN. The elastic energy

$$\mathcal{V}_{elastic}(\mathbf{x}) = \theta^2/2[B\,(\partial_x u_y - \partial_y u_x)^2 + \mu\,((\partial_x u_y + \partial_y u_x)^2 + (\partial_x u_x - \partial_y u_y)^2)]$$

describes the elastic penalty of deforming the layers. B and $\mu$ are the bulk and shear modulus respectively. Finally, the electrostatic energy is given by the coupling between an out-of-plane electric field and the local dipole in the polar domains: $\mathcal{V}_{elec}(\mathbf{x})= -\mathcal{E}p(\mathbf{x})$, where

$$p(\mathbf{x})=\sum_i p^e_n \phi^e_n(\mathbf{x})+p^o_n \phi^o_n(\mathbf{x})$$

is the local dipole moment as a function of relative stacking. The total energy was minimized to obtain the displacement field **u(x)** for fixed values of θ and $\mathcal{E}$, and the resulting stacking energy $\mathcal{V}_{tot}(\mathbf{x}+\mathbf{u}(\mathbf{x}))$ for each $\mathcal{E}$ is shown in Fig. 4(g) and Fig.19.

Fig. 19 shows the stacking energies after lattice relaxation for a twist angle of θ = 0.1°. At zero field, the relaxation reduces the area of the AA regions and increases the area of the AB/BA regions, leading to a triangular domain structure with sharp domain walls. When an electric field is applied, the AB and BA regions relax unevenly, leading to larger AB regions and smaller BA regions reducing the rotation symmetry to $C_3$. When the AB and BA domains are no longer equal in area, the polarization no longer averages to zero. In Fig. 20, we show the average spontaneous polarization as a function of the electrical field. It should be noted that the fields we explored theoretically are much larger than experimentally reachable fields.

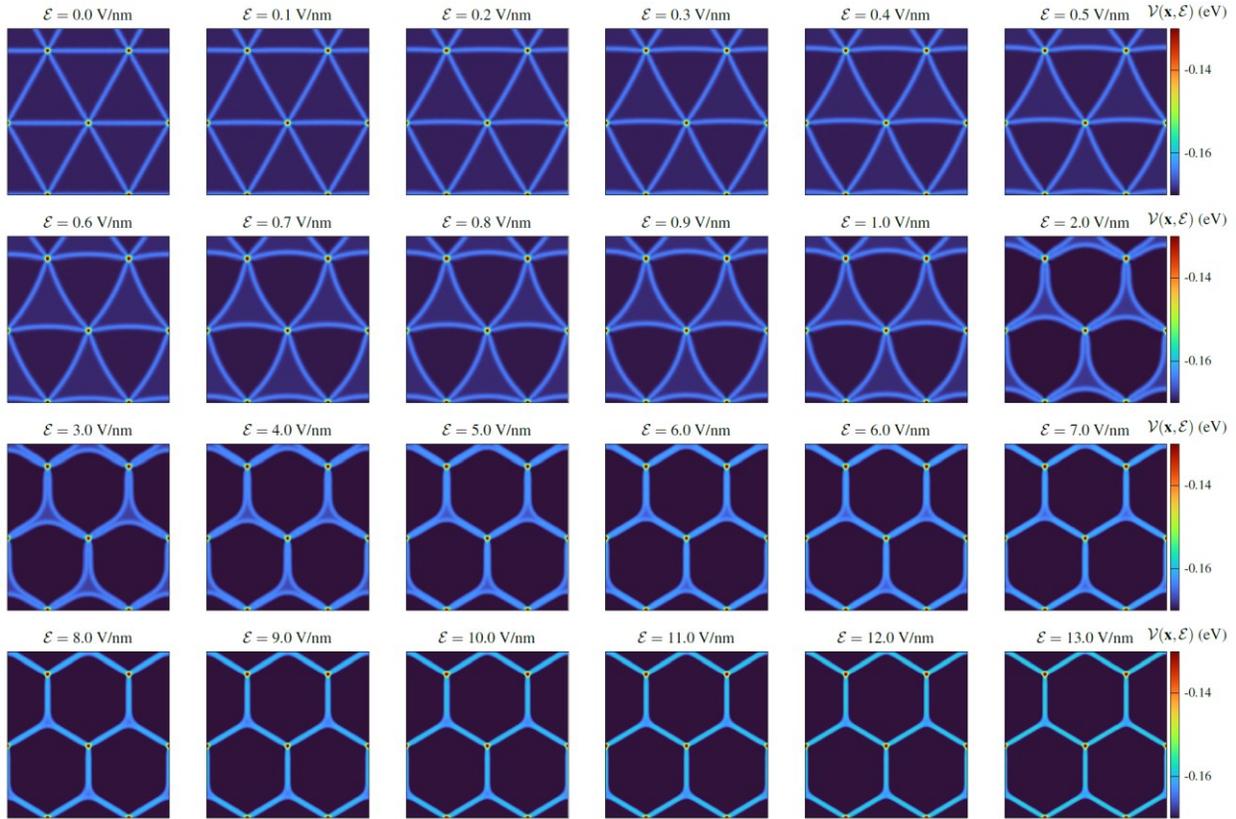

FIG. 19. Stacking energies in t-BN bilayer after lattice relaxation, as a function of position, for a twist angle of θ=0.1°.

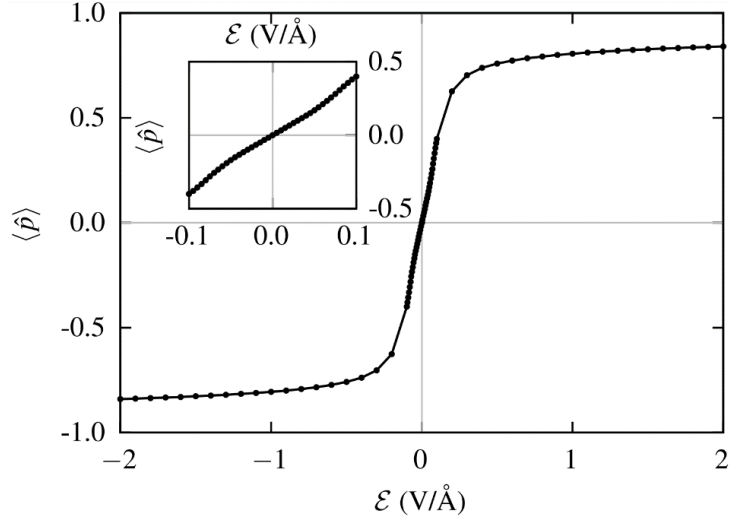

FIG. 20. Net polarization in t-BN bilayer after lattice relaxation, as a function of the applied electrical field, for a twist angle of $\theta$=0.1°. The inset is the zoom-in near zero field.